\documentclass[amsmath,amssymb,nofootinbib,prd]{revtex4}
\pdfoutput=1
\usepackage{graphicx}
\usepackage{amsmath}
\usepackage{amssymb}
\usepackage{amsfonts}

\begin{document}

\title{Beyond the plane-parallel and Newtonian approach:\\
Wide-angle redshift distortions and convergence in general relativity}

\author{Daniele Bertacca$^a$, Roy Maartens$^{a,b}$,
Alvise Raccanelli$^{c,d}$, Chris Clarkson$^{e}$\\~}

\affiliation{ $^a$Physics Department, University of the Western
Cape, Cape Town 7535, South Africa\\
$^b$Institute of Cosmology \& Gravitation, University of
Portsmouth,
Portsmouth PO1 3FX, UK \\
$^c$Jet Propulsion Laboratory, California Institute of Technology,
Pasadena CA 91109, US \\
$^d$California Institute of Technology,
Pasadena CA 91125, US\\
$^e$Centre for Astrophysics, Cosmology \& Gravitation, and, Department of Mathematics \& Applied Mathematics, University of Cape Town, Cape Town 7701, South Africa}

\begin{abstract}

We extend previous analyses of wide-angle correlations in the galaxy power spectrum in redshift space to include all general relativistic effects. These general relativistic corrections to the standard approach become important on large scales and at high redshifts, and they lead to new terms in the wide-angle correlations. We show that in principle the new terms can produce corrections of nearly 10\% on Gpc scales over the usual Newtonian approximation. General relativistic corrections will be important for future large-volume surveys such as SKA and Euclid, although the problem of cosmic variance will present a challenge in observing this.
\end{abstract}

\date{\today}

\maketitle

\section{Introduction}

Upcoming surveys of galaxies and HI will probe increasingly large
scales, approaching and even exceeding the Hubble scale at the
survey redshifts. On these cosmological scales, surveys can in
principle provide the best constraints on dark energy and modified
gravity models -- and will be able to test general relativity
itself. Furthermore, constraints on primordial non-Gaussianity
from  large-scale surveys of the matter distribution will be
competitive with CMB constraints. However, in order to realise the
potential of these surveys, we need to ensure that we are using a
correct analysis, i.e. a general relativistic analysis, on
cosmological scales.

There are two fundamental issues underlying the GR analysis.
\begin{itemize}
\item
We need to correctly identify the galaxy overdensity $\Delta$ that is {\em observed} on the past light cone. The
overdensity $\delta_g$ defined in different gauges
gives the same results on sub-Hubble scales, but leads to
different results on large scales -- and this remains true even if
we use gauge-invariant definitions of $\delta_g$. The observed
$\Delta$ is necessarily gauge-invariant, and is unique.

\item
We need to account for all the distortions arising from observing
on the past light cone, including redshift distortions (with all
GR effects included) and volume distortions.

\end{itemize}

These GR effects come in to the \emph{measured} 2-point correlation function in redshift space $\xi({\bf n_1},{\bf n_2};z_1,z_2)$. We provide a new representation of this function which is relatively simple to calculate, and which takes into account all GR and wide-angle contributions. Our fully general relativistic wide-angle formalism recovers and generalizes previous work in the Newtonian plane-parallel (flat-sky) \cite{Kaiser:1987qv, Hamilton:1997zq} and Newtonian wide-angle \cite{Szalay:1997cc, Bharadwaj:1998bq, Matsubara:1999du,Szapudi:2004gh, Papai:2008bd,Raccanelli:2010hk, Samushia:2011cs} cases.

\subsection*{Observed galaxy density perturbation}

The GR analysis of the matter power spectrum 
\cite{Yoo:2010jd,Yoo:2010ni,Bartolo:2010ec,Bonvin:2011bg,Challinor:2011bk,Bruni:2011ta,
Baldauf:2011bh,Jeong:2011as,Yoo:2011zc,Schmidt:2012ne,Jeong:2012nu,
Schmidt:2012nw} leads to corrections
on cosmological scales of the standard Newtonian analysis (which
is accurate on small scales). The
observed galaxy overdensity is a function of the observed
direction ${\bf n}$ and redshift $z$. It may be expressed in any
chosen gauge. We use synchronous-comoving gauge, in which
\begin{equation}\label{sync}
ds^2=a^2(\tau)\Big\{-d\tau^2+\Big[\big(1-2{\cal R}
\big)\delta_{ij}+2\partial_i \partial_j E\Big] dx^idx^j\Big\}.
\end{equation}
In $\Lambda$CDM, we have ${\cal R}'=0$
\cite{Ma:1995ey,Matarrese:1997ay} (a prime denotes
$\partial_\tau$).

We can write the observed overdensity at observed redshift $z$ and in the unit direction ${\bf n}$ as
\begin{equation}
\label{delta_g} \Delta( {\bf n},z) =  {\Delta}_s ( {\bf n},z)+
{\Delta}_\kappa( {\bf n},z)+ {\Delta}_I ( {\bf n},z).
 \end{equation}
Here $ {\Delta}_s $ is a local term (i.e. evaluated at the source)
which includes the galaxy density perturbation, the redshift
distortion and the change in volume entailed by the redshift
perturbation. $ {\Delta}_\kappa$ is the weak lensing convergence
integral along the line of sight, and $ {\Delta}_I$ is a time
delay integral along the line sight. In the gauge (\ref{sync}), we
have \cite{Jeong:2011as}
\begin{eqnarray}
\label{dels}{\Delta}_s &=&b \delta + \left[b_e -
\left(1+2\mathcal{Q}\right)+\frac{(1+ {z})}{H}\frac{dH}{dz}
-\frac{2}{ {\chi}}\left(1-\mathcal{Q}\right)\frac{(1+ {z})}
{H}\right]\left( \partial_\parallel E' + E'' \right) \nonumber\\
&&- \frac{(1+ {z})}{H} \partial^2_\parallel E' -\frac{2}{
{\chi}}\left(1-\mathcal{Q}\right)\left(\chi {\cal R}+E'\right)\;, \\
\label{delk}
 {\Delta}_\kappa &=& \left(1-\mathcal{Q}\right)\nabla^2_\perp
 \int_0^{ {\chi}} d\tilde\chi  \left( {\chi}-\tilde\chi\right)
 \frac{ {\chi}}{\tilde\chi}\left(E''-{\cal R} \right)\;\\ \label{deli}
 {\Delta}_I &=& -\frac{2}{ {\chi}}\left(1-\mathcal{Q}\right)
 \int_0^{ {\chi}} d\tilde\chi\left(E''-{\cal R} \right) \nonumber \\
&&+  \left[b_e - \left(1+2\mathcal{Q}\right)+\frac{(1+
{z})}{H}\frac{dH}{dz}-\frac{2}{{\chi}}\left(1-\mathcal{Q}\right)\frac{(1+ {z})}{H}\right]
\int_0^{ {\chi}} d\tilde\chi E'''.
\end{eqnarray}
Here $\chi(z)$ is the comoving distance, $b( {z})$ is the bias and
\begin{equation}
b_e ( {z})=-(1+ {z}) {d\ln  [n_g(1+z)^{-3}] \over d z},
\end{equation}
where $n_g$ is the background number density. (We have changed some of the notation in \cite{Jeong:2011as}.)
The directional
derivatives are defined as
\begin{eqnarray}
&&\partial_\parallel= { {{n}}}^j \partial_j,~~\partial_\parallel^i =
{ {{n}}}^i \partial_\parallel, ~~\partial^2_\parallel =
\partial_{\parallel i}\partial_\parallel^i =\partial_\parallel
\partial_\parallel,\\
&&\partial_\perp^i = (\delta^{ij} -  { {{n}}}^i {
{{n}}}^j)\partial_j= \partial^i -  { {{n}}}^i \partial_\parallel
,~~ \nabla^2_\perp = \partial_{\perp i}\partial_\perp^i =\nabla^2
- \partial_\parallel^2 - 2 {\chi}^{-1}\partial_\parallel.
\end{eqnarray}
The term
$\mathcal{Q}({z})$ encodes the magnification bias, which arises from the perturbation to the flux of
a galaxy, relative to a galaxy at the same observed redshift in
the unperturbed universe (see \cite{Jeong:2011as}).

The local term $\Delta_s$ contains the Newtonian local terms, and in addition some GR corrections. The line of sight term $\Delta_I$ is a pure GR correction. The lensing term $\Delta_\kappa$ is the same as in  the Newtonian analysis. Note that in $\nabla_\perp^2$ we do not drop the radial derivative term $-2\chi^{-1}\partial_\parallel$ which gives a negligible contribution to (\ref{delk}) on small scales, but should not be neglected on large scales.
 
We can relate the metric perturbations to the matter density
contrast in synchronous gauge, removing the residual gauge
ambiguity. Taking into account that $E''+aHE'-4\pi G \rho_m E =0$,
we get
\begin{eqnarray}
E'     &=&-\frac{H}{(1+z)}f\nabla^{-2}\delta, \\
E''    &=&-\frac{H^2}{(1+z)^2} \Big(\frac{3}{2}
\Omega_m-f\Big)\nabla^{-2}\delta, \\
E'''   &=&-3 \frac{H^3}{(1+z)^3}\Omega_m
\left(f-1\right)\nabla^{-2}\delta ,\\
{\cal R} &=&  \frac{H^2}{(1+z)^2} \Big(\frac{3}{2}
\Omega_m+f\Big) \nabla^{-2}\delta.
\end{eqnarray}
Here $\Omega_m(z)$ is the matter density and  $f(z)$ is the growth rate,
\begin{equation}
f={d\ln D \over d\ln a},~~~ \delta({\bf x},z)= \delta({\bf x},0){D(z) \over D(0)},
\end{equation}
where $D $ is the growing mode of $\delta$.

\section{Redshift-space correlation functions in general relativity}
\label{}

The correlation function can be decomposed in spherical harmonics
as \cite{Bonvin:2011bg}
\begin{equation}
\xi( {\bf n}_1, {\bf n}_2, z_1,z_2) = \langle \Delta( {\bf
n}_1,z_1) \Delta( {\bf n}_2,z_2)  \rangle=\sum_{\ell,m}
C_\ell(z_1,z_2) Y_{\ell m}({\bf n}_1)Y^*_{\ell m}({\bf n}_2). \label{xicl}
\end{equation}
However, this turns out to be computationally expensive to
implement for wide-angle correlations, and we follow the
alternative decomposition used in previous analyses based on a
Newtonian approach \cite{Szapudi:2004gh, Papai:2008bd,
Raccanelli:2010hk, Samushia:2011cs}.
(The formula for the correlation function in redshift space,
including wide-angle effects, was first derived in
\cite{Szalay:1997cc,Matsubara:1999du}, without GR corrections.) This alternative expands
the redshift space correlation function using tripolar spherical harmonics, with the basis functions
\begin{eqnarray}
&& S_{\ell_1\ell_2 L}( { {{\bf n}}}_1,  { {{\bf n}}}_2, { {{\bf
n}}}_{12}) =\left[ {(4\pi)^3 \over (2\ell_1+1) (2\ell_2+1)
(2L+1)} \right]^{1/2}\! \sum_{m_1,m_2,M} \left(
\begin{array} {ccc} \ell_1 & \ell_2 & L \\ m_1 & m_2 & M
\end{array} \right)
Y_{\ell_1 m_1}( { {{\bf n}}}_1) Y_{\ell_2 m_2}( { {{\bf n}}}_2) Y_{L
M}( { {{\bf n}}}_{12}),  \nonumber \\ &&-\ell_1 \leq m_1 \leq \ell_1\;, \quad -\ell_2 \leq m_2 \leq \ell_2\;, \quad -L \leq M \leq L\,. \label{tripo}
\end{eqnarray}

Here
\[
\left(
\begin{array} {ccc} \ell_1 & \ell_2 & \ell_3 \\ m_1 & m_2 & m_3
\end{array} \right)
\]
is the Wigner 3$j$ symbol. Because our triangles are closed, only two independent $\ell$'s appear. The expansion for $\xi$ may then be written in redshift space as a sum over these functions with $L$ in the range $|\ell_1-\ell_2|\leq L \leq \ell_1+\ell_2$ (see \cite{Szapudi:2004gh} for details). In the conventional expansion \eqref{xicl} of $\xi$ into $C_\ell$'s, the $\ell$ represents the angular momentum of the correlation function, and is summed to infinity. By contrast, in the tripolar expansion, the $\ell$'s are representative of the radial derivatives $\partial_\|$ in the two directions ${\bf n}_1$ and ${\bf n}_2$. Consequently, the sums involved in the tripolar expansion are no longer infinite, making explicit computations in redshift space comparatively simple.

\begin{center}
\begin{figure*}[htb!]
\includegraphics[width=0.65\columnwidth]{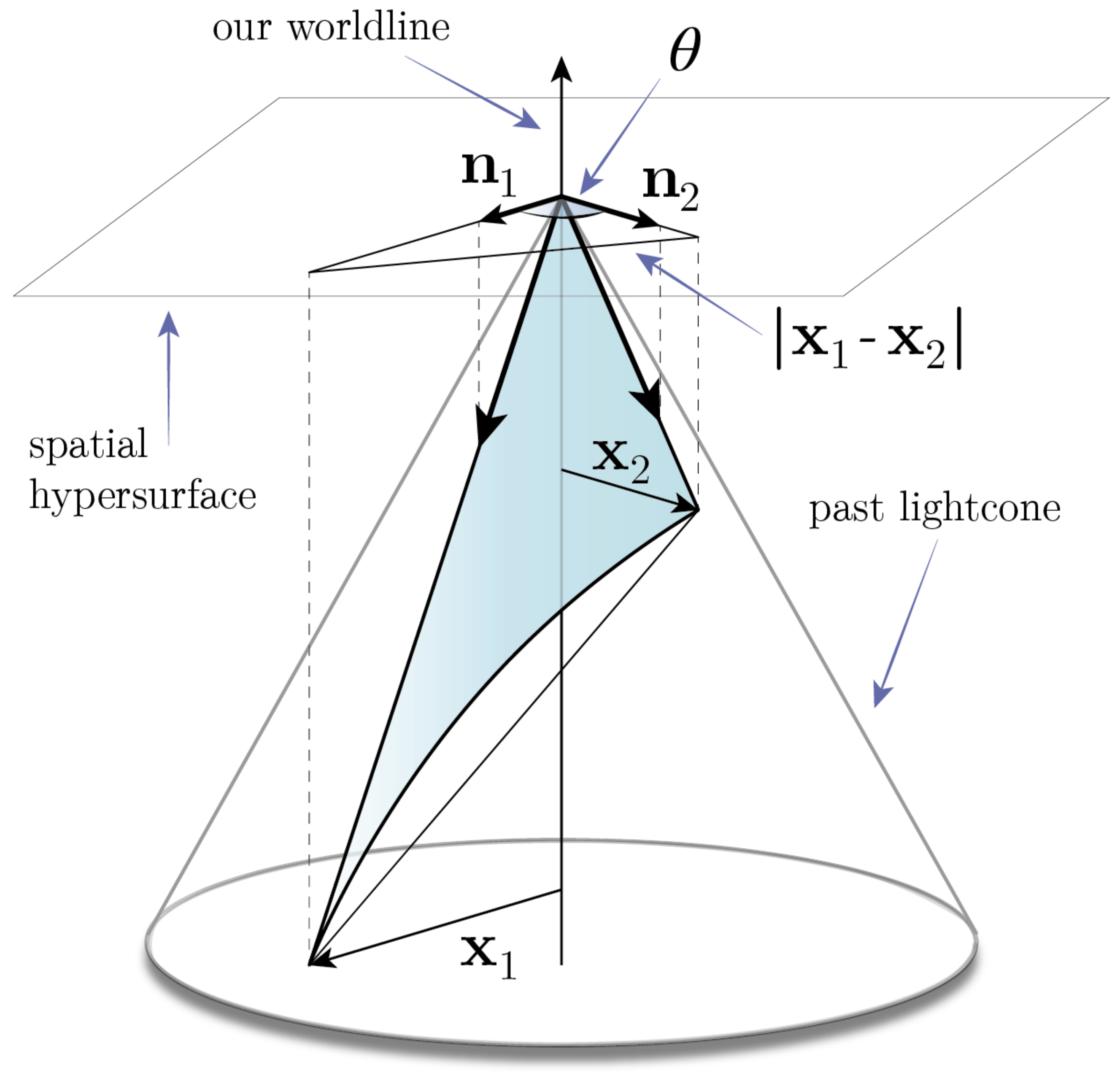}
\caption{Positions of a pair of galaxies on the lightcone.}\label{lcone}
\end{figure*}
\end{center}

The positions of two galaxies and their separation are given by (see Fig. \ref{lcone})
\begin{equation}
{\bf x}_1= {\chi}_1  { {\bf n}}_1,~~ {\bf x}_2= {\chi}_2 { {\bf
n}}_2,~~ {\bf x}_{12}= {\bf x}_1- {\bf x}_2 \equiv \chi_{12} {\bf
n}_{12}.
\end{equation}
The overdensity in the form (\ref{delta_g}) leads to the cross
correlation functions in redshift space:
\begin{eqnarray}
\label{xi} \xi_{AB}( {\bf x}_1, {\bf x}_2) = \langle
\Delta_A( {\bf x}_1) \Delta_B( {\bf x}_2)  \rangle = \xi_{BA}({\bf x}_2, {\bf x}_1), ~~  A,B= s,
\kappa,I. \label{xiab}
\end{eqnarray}

First, we define the spherical transform of the matter
overdensity \cite{Szalay:1997cc}
\begin{equation}
\label{A-tensor} \mathcal{A}^n_\ell ({\bf x},z)= \int
\frac{d^3k}{(2\pi)^3} (ik)^{-n}  \, \mathcal{P}_\ell ( {{\bf n}}
\cdot  {\hat{\bf k}})\exp{\left(i{\bf k} \cdot {\bf x}\right)}\;
\delta({\bf k},z),~~~ {\bf x}=\chi {\bf n},
\end{equation}
where $\mathcal{P}_\ell$ is a Legendre polynomial. Then by \eqref{dels}, ${\Delta}_s$ can be be decomposed as
\begin{eqnarray}
{\Delta_s \over b} = \left(1+\frac{1}{3}\beta\right)
\mathcal{A}^0_0+ \gamma \mathcal{A}^2_0  + \frac{\beta\alpha}{\chi}
\mathcal{A}^1_1  +\frac{2}{3}\beta \mathcal{A}^0_2,\label{deltas}
 \end{eqnarray}
where 
\begin{eqnarray}
\label{alpha}
\alpha(z)&=& - \chi(z) \frac{H(z)}{(1+z)} \left[b_e(z) -
1-2\mathcal{Q}(z) +\frac{3}{2}\Omega_m (z)-
\frac{2}{\chi(z)}\big[1-\mathcal{Q}(z)\big]\frac{(1+z)}{H(z)}\right] \;, \\
\label{beta}
\beta(z)  &=& \frac{f(z)}{b(z)}\;, \\
\gamma(z) &=& \frac{H(z)}{(1+z)} \left\{\frac{H(z)}{(1+z)}
\left[\beta(z) -\frac{3}{2}\frac{\Omega_m (z)}{b(z)}
\right]b_e(z) + \frac{3}{2}\frac{H(z)}{(1+z)} \beta(z) \big[
\Omega_m (z) - 2 \big]
\right. \nonumber \\
& & \left. -\frac{3}{2} \frac{H(z)}{(1+z)}  \frac{\Omega_m
(z)}{b(z)} \left[1-4\mathcal{Q}(z) + \frac{3}{2}\Omega_m
(z)\right] +\frac{3}{\chi(z)}
\big[1-\mathcal{Q}(z)\big]\frac{\Omega_m (z)}{b(z)} \right\} .
\label{gamma}
\end{eqnarray}
Here $\alpha$ is a generalization of the Newtonian expression, $\beta$ has the same form as in the Newtonian analysis and $\gamma$ is a new term arising from GR corrections: see Fig. \ref{abgms} below.

In order to decompose $\xi_{ss}$, we need the correlator of
(\ref{A-tensor}):
\begin{eqnarray}
\langle \mathcal{A}^{n_1}_{\ell_1} ( {\bf x}_1,
{z}_1)\mathcal{A}^{n_2}_{\ell_2} ( {\bf x}_2, {z}_2) \rangle =
(-1)^{\ell_2} \int  \frac{d^3k}{(2\pi)^3} (ik)^{-(n_1+n_2)}
\mathcal{P}_{\ell_1}( {\hat{\bf k}} \cdot   { {{\bf n}}}_1)
\mathcal{P}_{\ell_2}( {\hat{\bf k}} \cdot   { {{\bf n}}}_2)
\exp{\left(i{\bf k} \cdot  {{\bf x}}_{12}\right)} \; P_\delta(k;
{z}_1, {z}_2)\;,
\end{eqnarray}
where $P_\delta(k; {z}_1, {z}_2)$ is defined by
\begin{equation}
\langle \delta({\bf k}_1, {z}_1) \delta({\bf k}_2, {z}_2) \rangle
= (2\pi)^3 \delta^3_D ({\bf k}_1+ {\bf k}_2) P_\delta(k_1; {z}_1,
{z}_2)\;.
\end{equation}
In terms of the primordial power spectrum and the transfer
function, we have
\begin{equation}
P_\delta(k; {z}_1, {z}_2)=P_{\rm prim}(k)T^2(k){D(z_1)D(z_2) \over
D^2(0)}\,.
\end{equation}
Expanding $ \mathcal{P}_{\ell}$ and $\exp{\left(i{\bf k} \cdot
{\bf x}\right)}$ in spherical harmonics and applying the Gaunt
integral \cite{Varshalovich:1988ye}, we obtain
\begin{eqnarray}
&& \langle \mathcal{A}^{n_1}_{\ell_1} ( {\bf x}_1,
{z}_1)\mathcal{A}^{n_2}_{\ell_2} ( {\bf x}_2, {z}_2)
\rangle=\sum_{L} (-1)^{\ell_2} \; i^{L-n_1-n_2} \left( \begin{array}
{ccc} \ell_1 & \ell_2 & L \\ 0 & 0 & 0
\end{array} \right) S_{\ell_1\ell_2L}(
{ {{\bf n}}}_1,  { {{\bf n}}}_2, { {{\bf n}}}_{12})\,
\xi_L^{n_1+n_2}(\chi_{12};z_1, z_2) , \nonumber\\ && |\ell_1-\ell_2|
\leq L \leq \ell_1+\ell_2. \label{aa}
\end{eqnarray}
This is an expansion in the
tripolar basis functions \eqref{tripo},
with coefficients
\begin{equation}\label{xiLn}
\xi_L^{n}(\chi; z_1, z_2) = \int \frac{dk}{2\pi^2} k^{2-n}
j_L(\chi k) \, P_\delta(k; z_1, z_2)\;.
\end{equation}

Finally, we arrive at the tripolar decomposition of $\xi_{ss}$ in the most general (GR wide-angle) case:
\begin{equation}
{\xi}_{ss}( {\bf x}_1, {\bf x}_2) =b(z_1)b(z_2)
\sum_{\ell_1,\ell_2,L,n} B_{ss \; n}^{\phantom{ss
\;}\ell_1\ell_2L}( {\chi}_1, {\chi}_2)\, S_{\ell_1\ell_2L}( {
{{\bf n}}}_1, { {{\bf n}}}_2,  { {{\bf n}}}_{12})\,
\xi_L^{n}(\chi_{12}; z_1, z_2). \label{eq:xi_ss}
\end{equation}
By \eqref{deltas}, $\ell_i,n_i=0,1,2$ so that $n\equiv n_1+n_2=0,1,2,3,4$; by \eqref{aa} we also have $L=0,1,2,3,4$.
The $B$ coefficients for $\xi_{ss}$ and the other $\xi_{AB}$ (see
below) are given in Appendix \ref{bco}. These coefficients involve the functions \eqref{alpha} and \eqref{gamma} that contain GR corrections. In order to compare the GR and Newtonian cases, we set $\mathcal{Q}=0$ for simplicity.
We rewrite (\ref{alpha}) as
\begin{eqnarray}\label{alpha2}
\frac{\alpha(z)}{\chi(z)} =
-  \frac{H(z)}{(1+z)} \left[\frac{3}{2}\Omega_m (z)-1\right] + \frac{d
\ln{N_g}}{d \chi}+ \frac{2}{\chi},
\end{eqnarray}
where $N_g$ is the comoving galaxy number density. Then for  small redshift (i.e.
$\chi \to 0$), we recover the Newtonian limit of $\alpha$  \cite{Hamilton:1997zq}:
\begin{equation}
\frac{\alpha}{\chi} \to \frac{\alpha_N}{\chi}=  \frac{d \ln{N_g}}{d \chi}+ \frac{2}{\chi}.
\end{equation}
In the same limit, \eqref{gamma} shows that
\begin{eqnarray}
\gamma(z)  &\to & \frac{3\Omega_m(z)}{2b(z)}\frac{H(z)}{(1+z)} \left\{\left[1-\frac{2f(z)}
{3\Omega_m(z)}\right] \frac{d \ln{n_g}}{d \chi} + \frac{2}{\chi} \right\},
\label{gamma3}
\end{eqnarray}
which is nonzero. However, to recover the Newtonian limit, we have to take into account (\ref{xiLn}) when
$\chi_{12}\to 0$. Then we find:
\begin{equation}
\gamma \xi^n_L \to 0,
\end{equation}
so that $\gamma$ drops out of $\xi_{ss}$ in the Newtonian approximation.
\begin{center}
\begin{figure*}[htb!]
\includegraphics[width=0.47\columnwidth]{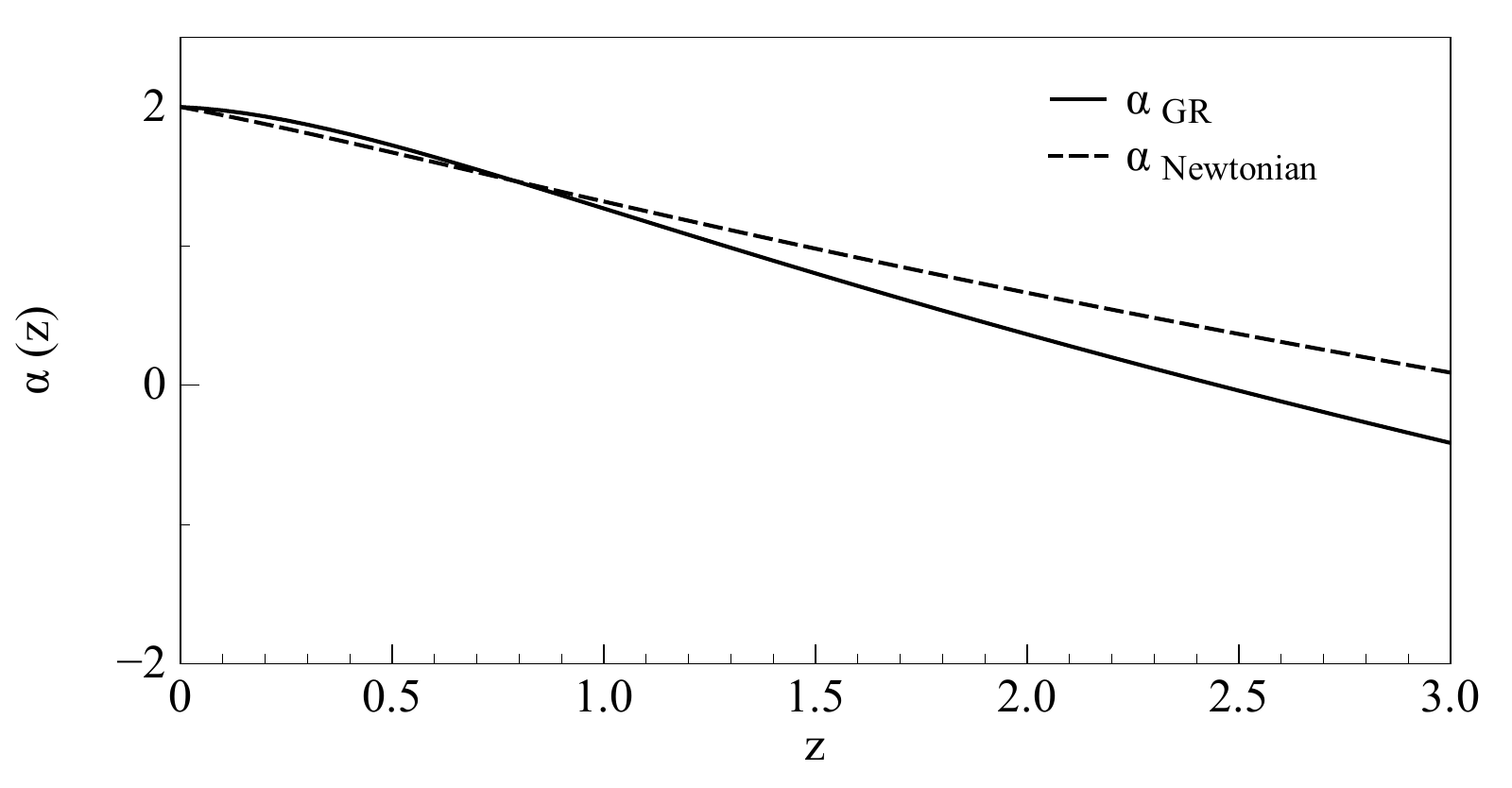}~~~~
\includegraphics[width=0.47\columnwidth]{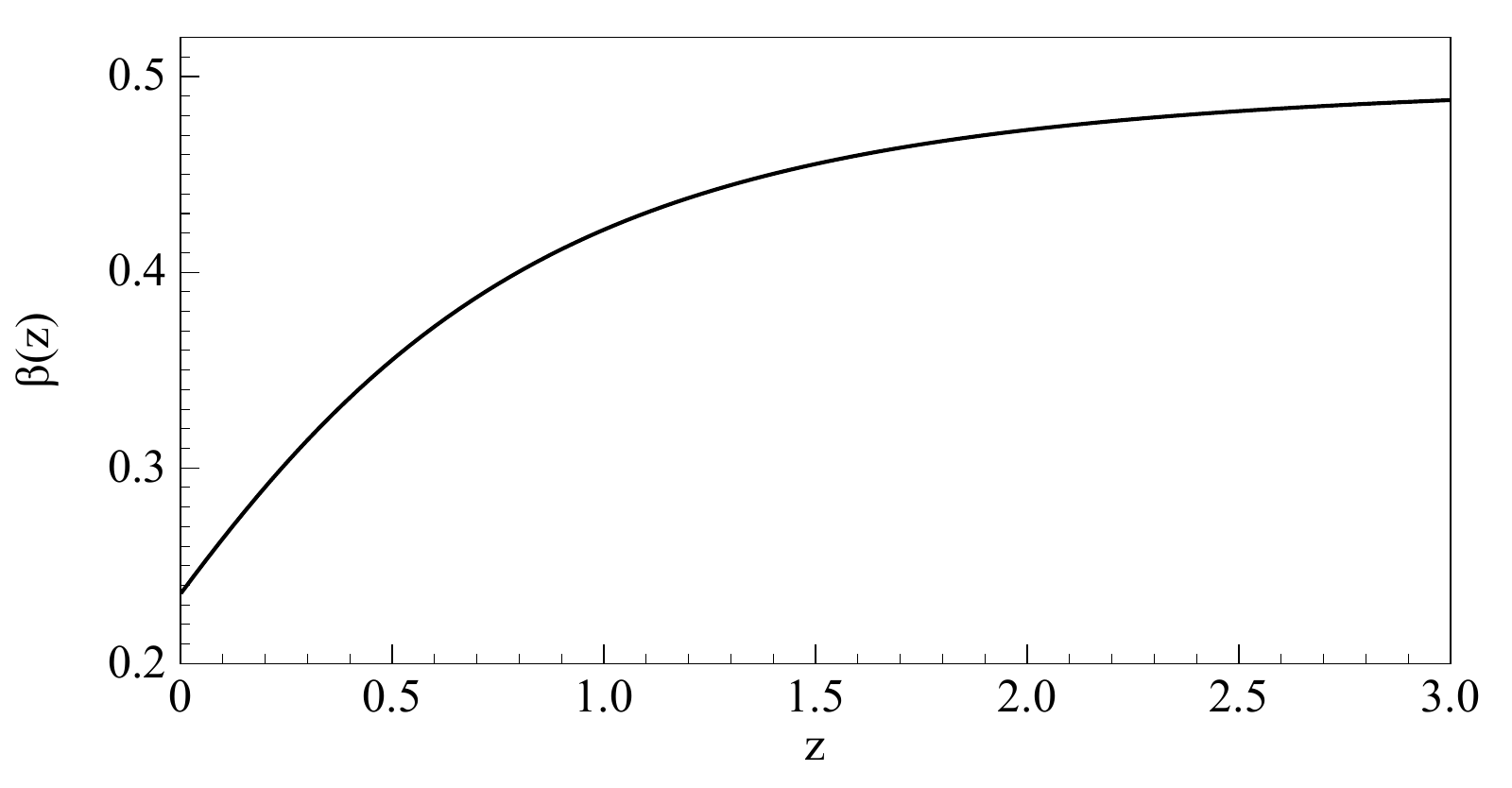}\\
\includegraphics[width=0.47\columnwidth]{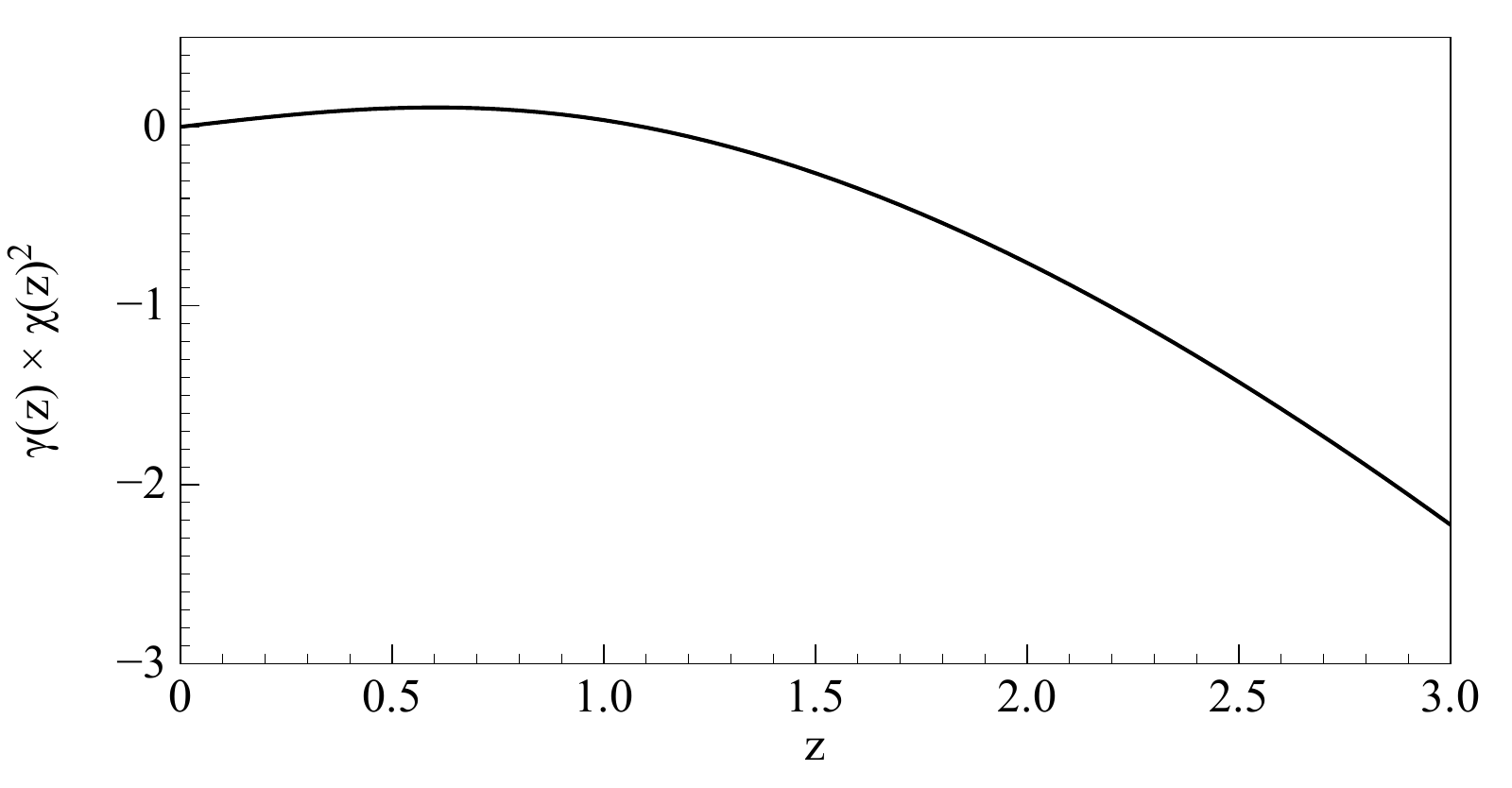}
\caption{The functions $\alpha,\beta,\gamma$ in \eqref{deltas}, assuming a concordance model, and with ${\cal Q}=0$,  $b= 2$ and $n_g$ as defined in~\cite{Jeong:2011as}. $\beta$ is unchanged by GR corrections, whereas $\gamma$ is not present in the Newtonian analysis. }\label{abgms}
\end{figure*}
\end{center}

\begin{center}
\begin{figure*}[htb!]
\includegraphics[width=0.47\columnwidth]{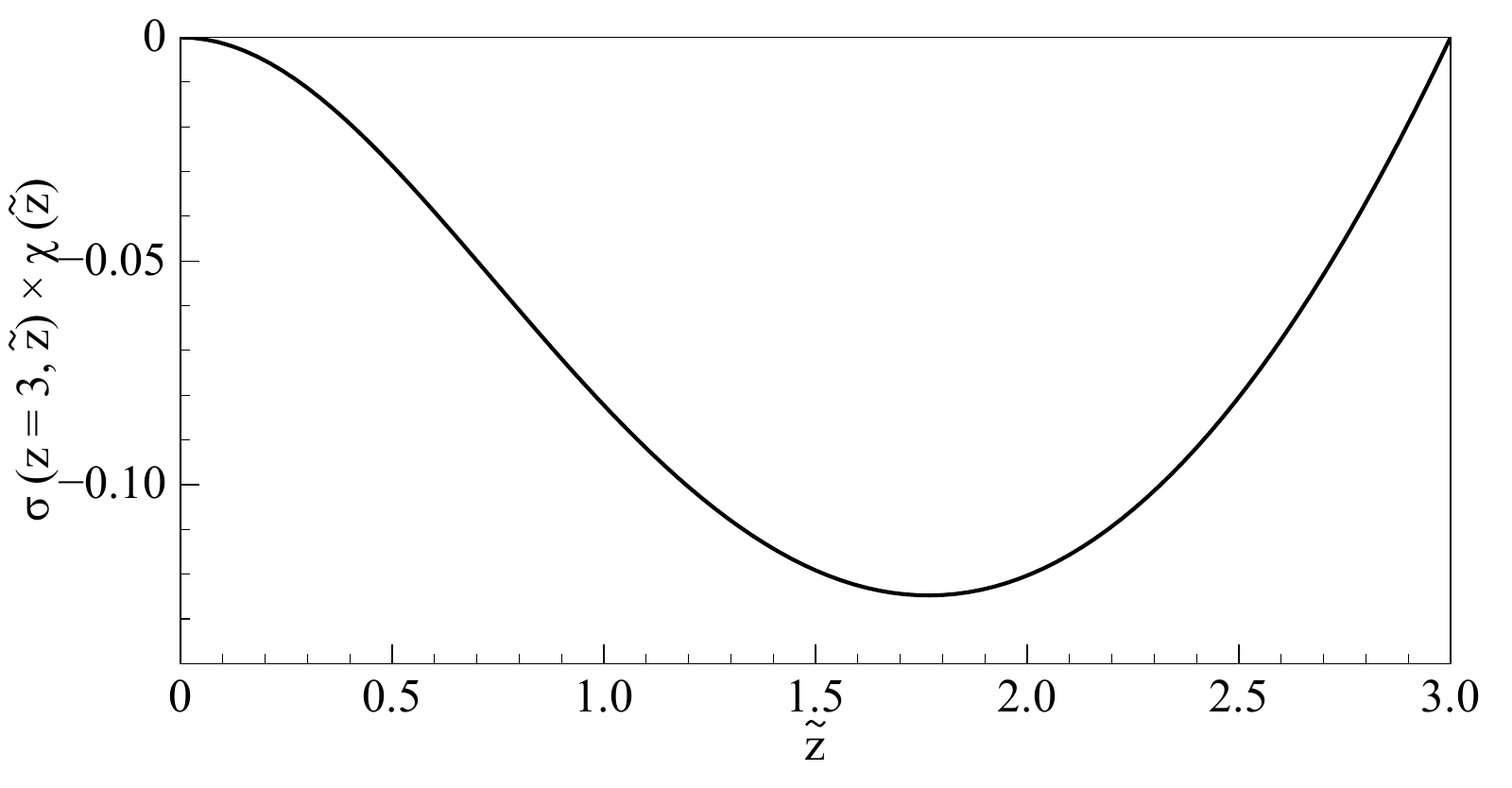}~~~~
\includegraphics[width=0.47\columnwidth]{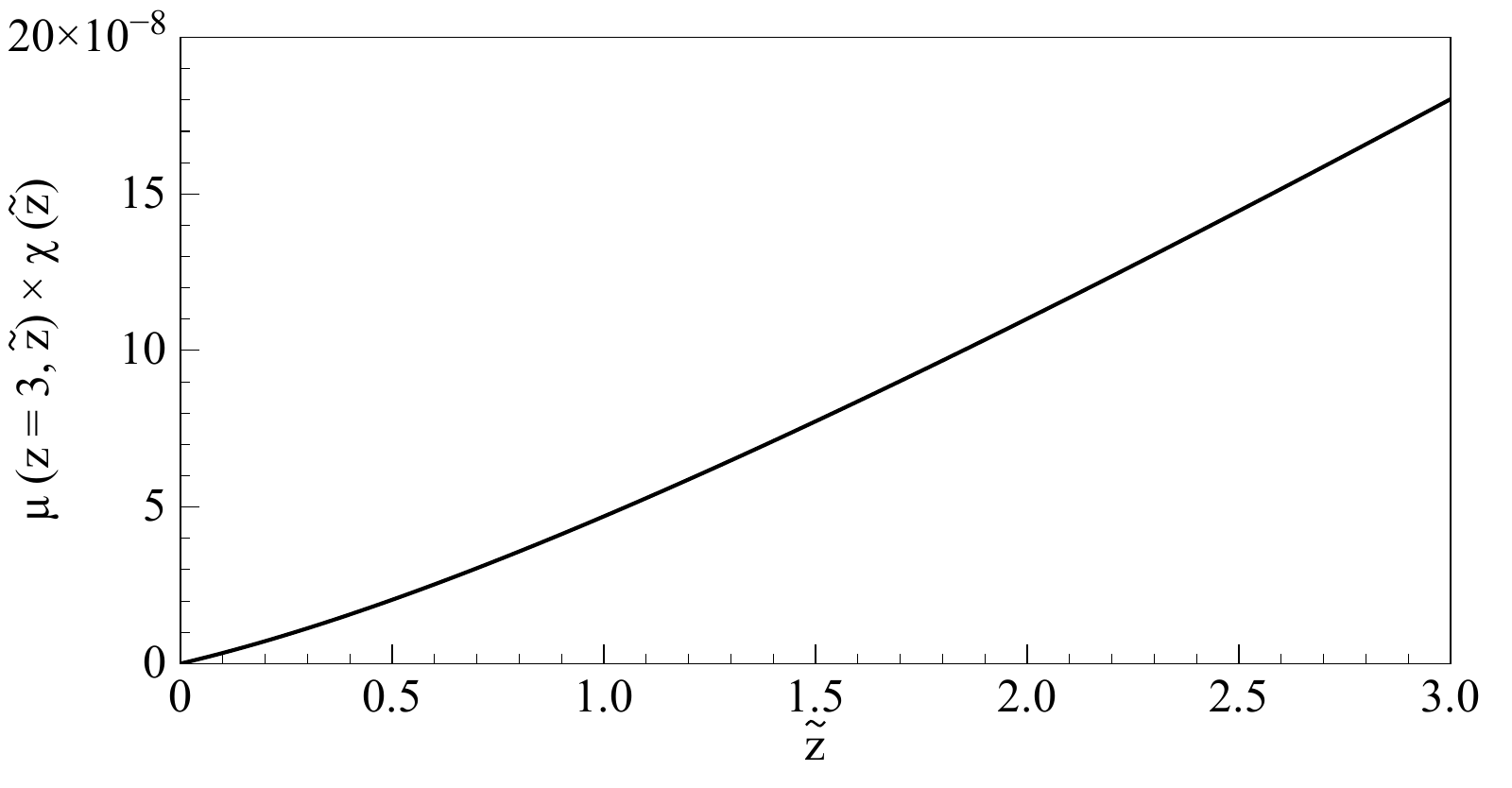}
\caption{The functions  $\sigma(3,\tilde z)$ and $\mu(3,\tilde z)$, assuming a concordance model, and with ${\cal Q}=0$,  $b= 2$ and  $n_g$ as defined in~\cite{Jeong:2011as}. $\sigma$ is unchanged by GR corrections, whereas $\mu$ is not present in the Newtonian analysis. }\label{sigmu}
\end{figure*}
\end{center}

The remaining $\xi_{AB}$ all involve integrals along the lines
of sight. The spherical transforms of $\Delta_\kappa, \Delta_I$ are
\begin{eqnarray}
\Delta_\kappa ({\bf n}, z)&=& b(z)\int^{\chi}{d\tilde\chi} \; \sigma(z,\tilde{z})\left[\mathcal{A}^0_0(\tilde{\bf x},\tilde{z})- \mathcal{A}^0_2(\tilde{\bf x},\tilde{z})-\frac{3}{\tilde{\chi}}\mathcal{A}^1_1(\tilde{\bf x},\tilde{z})\right], \label{dkexp} \\
\Delta_I ({\bf n}, z)&=& b(z)\int^{\chi}{d\tilde\chi} \; \mu(z,\tilde{z})\mathcal{A}^2_0(\tilde{\bf x},\tilde{z}), \label{diexp}
\end{eqnarray}
where 
\begin{eqnarray}
\sigma( z,\tilde z) &\equiv & -2\frac{H^2(\tilde{z})}{(1+\tilde{z})^2} \frac{\left({\chi}-\tilde\chi\right)\tilde\chi}{ {\chi}}\frac{\big[1-\mathcal{Q}(z)\big]}{b( {z})}   \Omega_m (\tilde z) , \label{sigma}\\
\mu( z,\tilde z) &\equiv & 3
\frac{H^2(\tilde{z})}{(1+\tilde{z})^2}\frac{\Omega_m
(\tilde{z})}{b( {z})}\left\{\frac{2}{ {\chi}} \big[1-\mathcal{Q}(z)
\big] \right. \nonumber \\
&& \left.  - \frac{H(\tilde{z})}{(1+\tilde{z})}
\big[f(\tilde{z})-1\big] \left[ b_e( {z}) - \big[1+2\mathcal{Q}(z)\big] +\frac{3}{2} \Omega_m ( {z})- \frac{2}{
{\chi}}\big[1-\mathcal{Q}( z )\big] \frac{(1+ {z})}{H(
{z})}\right]  \right\}, \label{mu}
\end{eqnarray}
and $\chi=\chi(z), \tilde\chi=\chi(\tilde z)$. These functions are illustrated in Fig. \ref{sigmu}: $\sigma$ has the same form as in Newtonian analysis and $\mu$ corresponds to GR terms that vanish in the Newtonian limit.

For the lensing-lensing correlation, we find
\begin{equation}
{\xi}_{\kappa \kappa}( {\bf x}_1, {\bf x}_2)
=b(z_1)b(z_2) \int^{ {\chi}_1,\chi_2}
{d\tilde\chi_1}{d\tilde\chi_2} \sum_{\ell_1,\ell_2,L,n} B_{\kappa
\kappa \; n}^{\phantom{\kappa \kappa \;}\ell_1\ell_2L}(
{\chi}_1,\tilde\chi_1; {\chi}_2, \tilde\chi_2)\,
S_{\ell_1\ell_2L}( { {{\bf n}}}_1, { {{\bf n}}}_2, {\tilde{\bf
n}}_{12})\, \xi_L^{n}(\tilde\chi_{12}; \tilde{z}_1, \tilde{z}_2),
\label{eq:xi_kk}
\end{equation}
and for the ${II}$ correlation
\begin{equation}
{\xi}_{II}( {\bf x}_1, {\bf x}_2) =b(z_1)b(z_2)
\int^{ {\chi}_1,\chi_2} {d\tilde\chi_1}{d\tilde\chi_2}
\sum_{\ell_1,\ell_2,L,n} B_{II \; n}^{\phantom{\kappa
\kappa \;}\ell_1\ell_2L}( {\chi}_1,\tilde\chi_1; {\chi}_2,
\tilde\chi_2)\, S_{\ell_1\ell_2L}( { {{\bf n}}}_1, { {{\bf n}}}_2,
{\tilde{\bf n}}_{12})\, \xi_L^{n}(\tilde\chi_{12}; \tilde{z}_1,
\tilde{z}_2). \label{eq:xi_II}
\end{equation}
The integration variables ${\tilde\chi_{12}},{\tilde{\bf n}}_{12}$
are given by
\begin{eqnarray}
{\tilde\chi_{12}}{\tilde{\bf n}}_{12}&=& {\chi}_{12} { {{\bf
n}}}_{12}+ \left(\tilde\chi_1 - {\chi}_1\right) { {{\bf
n}}}_1-\left(\tilde\chi_2 -
{\chi}_2\right) { {{\bf n}}}_2  ,\label{kk-ii1}  \\
\tilde\chi^2_{12}&=& { \tilde\chi_1^2+ \tilde\chi_2^2 + \frac{
\tilde\chi_1\tilde\chi_2}{ {\chi}_1 {\chi}_2}\left[
{\chi}_{12}^2-\left( {\tilde\chi}_1^2+
{\tilde\chi}_2^2\right)\right]}.\label{kk-ii2}
\end{eqnarray}

Similarly, we find:
\begin{eqnarray}
{\xi}_{sI}( {\bf x}_1, {\bf x}_2) &=& b(z_1)b(z_2)
\int^{ {\chi}_2}{d\tilde\chi_2} \sum_{\ell_1,\ell_2,L,n} B_{sI \;
n}^{\phantom{sI \;}\ell_1\ell_2L}( {\chi}_1; {\chi}_2,
\tilde\chi_2) S_{\ell_1\ell_2L}( { {{\bf n}}}_1, { {{\bf n}}}_2,
{{\bf n}}_{1 \tilde{2}}) \, \xi_L^{n}(\chi_{1 \tilde{2}}; {z}_1,
\tilde{z}_2)  , \label{eq:xi_sI}\\
{\xi}_{s\kappa}( {\bf x}_1, {\bf x}_2) &=&
b(z_1)b(z_2) \int^{ {\chi}_2}{d\tilde\chi_2}
\sum_{\ell_1,\ell_2,L,n} B_{s\kappa \; n}^{\phantom{sI
\;}\ell_1\ell_2L}( {\chi}_1; {\chi}_2, \tilde\chi_2)
S_{\ell_1\ell_2L}( { {{\bf n}}}_1, { {{\bf n}}}_2, {{\bf n}}_{1
\tilde{2}}) \, \xi_L^{n}(\chi_{1 \tilde{2}}; {z}_1, \tilde{z}_2) ,
\label{eq:xi_sk}\\
{\xi}_{\kappa I}( {\bf x}_1, {\bf x}_2) &=&
b(z_1)b(z_2) \int^{ {\chi}_1,\chi_2}
{d\tilde\chi_1}{d\tilde\chi_2} \sum_{\ell_1,\ell_2,L,n} B_{\kappa
I \; n}^{\phantom{\kappa \kappa \;}\ell_1\ell_2L}(
{\chi}_1,\tilde\chi_1; {\chi}_2, \tilde\chi_2)\,
S_{\ell_1\ell_2L}( { {{\bf n}}}_1, { {{\bf n}}}_2, {\tilde{\bf
n}}_{12})\, \xi_L^{n}(\tilde\chi_{12}; \tilde{z}_1, \tilde{z}_2),
\label{eq:xi_kI}
\end{eqnarray}
where
\begin{eqnarray}
\chi_{1 \tilde{2}}{{\bf n}}_{ {1}\tilde 2}&=& {\left(
{\chi}_2-\tilde\chi_2 \right) { {{\bf n}}}_2 +  {\chi}_{12} {
{{\bf n}}}_{12}},\\
\chi_{1 \tilde{2}}^2 &=&{  {\chi}_1^2+ \tilde\chi_2^2 + \frac{
\tilde\chi_2}{ {\chi}_2}\left[ {\chi}_{12}^2-\left( {\chi}_1^2+
{\chi}_2^2\right)\right]}.
\end{eqnarray}

Note that the remaining $\xi_{AB}$ follow from the symmetry in \eqref{xiab}. (See Appendix \ref{bco} for the explicit expressions.)


\section{Computing the general relativistic correlations}

Here we consider the GR wide-angle correlation function $\xi_{ss}({\bf x}_1,{\bf x}_2)$ in various limits and then in the general case. For simplicity, and since it does not affect the comparisons, we assume  $\mathcal{Q}=0$.

\subsection*{GR small-angle (plane-parallel) limit}

For small angle and small galaxy separation, we have the plane-parallel or flat-sky limit, i.e. ${\bf n}_1$ and ${\bf n}_2$ are almost parallel \cite{Kaiser:1987qv, Hamilton:1997zq}. 

Using the properties of the Wigner coefficients and spherical functions, we get \cite{Szapudi:2004gh}
\begin{equation}
S_{\ell_1\ell_2 L}({\bf n}_1,{\bf n}_1,{\bf n}_{11})
=  \left( \begin{array} {ccc} \ell_1 & \ell_2  &L  \\ 0 & 0 & 0
\end{array} \right) \mathcal{P}_{L}({\bf n}_1 \cdot {\bf n}_{11}),
~~\mbox{where}~~ {\bf n}_1 \simeq {\bf n}_2,~ {\bf n}_{11} \equiv {\bf n}_{12}.
\end{equation}
With $  {z}_1\simeq  {z}_2$ ($ {\chi}_1\sim {\chi}_2$), $\chi_{11}$ is given by $ {\bf x}_{11}= {\chi}_{11} { {{\bf n}}}_{11}$.
From (\ref{eq:xi_ss}) we obtain the expansion of the redshift space correlation function in monopole, quadrupole and hexadecapole terms:
\begin{eqnarray}
{{\xi}_{ss}( {\bf x}_1, {\bf x}_1) \over b_1^{2}} &=&  \bigg\{ \left(1+\frac{2}{3}\beta_1 + \frac{1}{5}\beta_1^2 \right) \xi_0^{0}(  {\chi}_{11}; {z}_1,{z}_1)
   -   \bigg[ 2 \left(1+\frac{1}{3}\beta_1 \right) \gamma_1  -\frac{\beta_1^2 \alpha_1^2}{3\chi_1^2} \bigg] \xi_0^{2}(  {\chi}_{11}; {z}_1,{z}_1) \nonumber \\
   &&~+   \gamma_1^2  \xi_0^{4}( {\chi}_{11}; {z}_1,{z}_1) \bigg\} \mathcal{P}_{0}( { {{\bf n}}}_1 \cdot  { {{\bf n}}}_{11})
      +  \left[- 4 \beta_1 \left( \frac{1}{3}+ \frac{1}{7}\beta_1\right)\xi_2^{0}( {\chi}_{11}; {z}_1,{z}_1) \right.\nonumber \\
   &&~+ \frac{2}{3} \beta_1 \left( 2 \gamma_1- \frac{\beta_1\alpha_1^2}{\chi_1^2}\right)\xi_2^{2}( {\chi}_{11}; {z}_1,{z}_1)\bigg] \mathcal{P}_{2}( { {{\bf n}}}_1 \cdot  { {{\bf n}}}_{11})
   + \frac{8}{35} \beta_1^2 \xi_4^{0}( {\chi}_{11}; {z}_1,{z}_1) \mathcal{P}_{4}( { {{\bf n}}}_1 \cdot  { {{\bf n}}}_{11}). \label{ppxiss}
 \end{eqnarray}
(This is consistent with \cite{Yoo:2010ni, Jeong:2011as}.) In  \eqref{ppxiss}, we can divide the terms into 3 categories: those including only $\beta$, those with $\alpha$, and those with $\gamma$.
The terms with $\beta$ are the ones considered in the standard Kaiser analysis, and if we consider only them we will recover the classical results (see e.g. \cite{Hamilton:1997zq}).
The terms including $\alpha$ are responsible for the ``mode-coupling'' effects, and they arise in the jacobian relating real- to redshift-space (for details, in Newtonian analysis, see e.g. \cite{Szalay:1997cc, Szapudi:2004gh, Papai:2008bd, Raccanelli:2010hk}). These terms are usually ignored when using the plane-parallel approximation. Note that they vanish if the comoving radial distribution of galaxies is constant.
Finally, the terms with $\gamma$ are the GR corrections.

\begin{center}
\begin{figure*}[htb!]
\includegraphics[width=0.47\columnwidth]{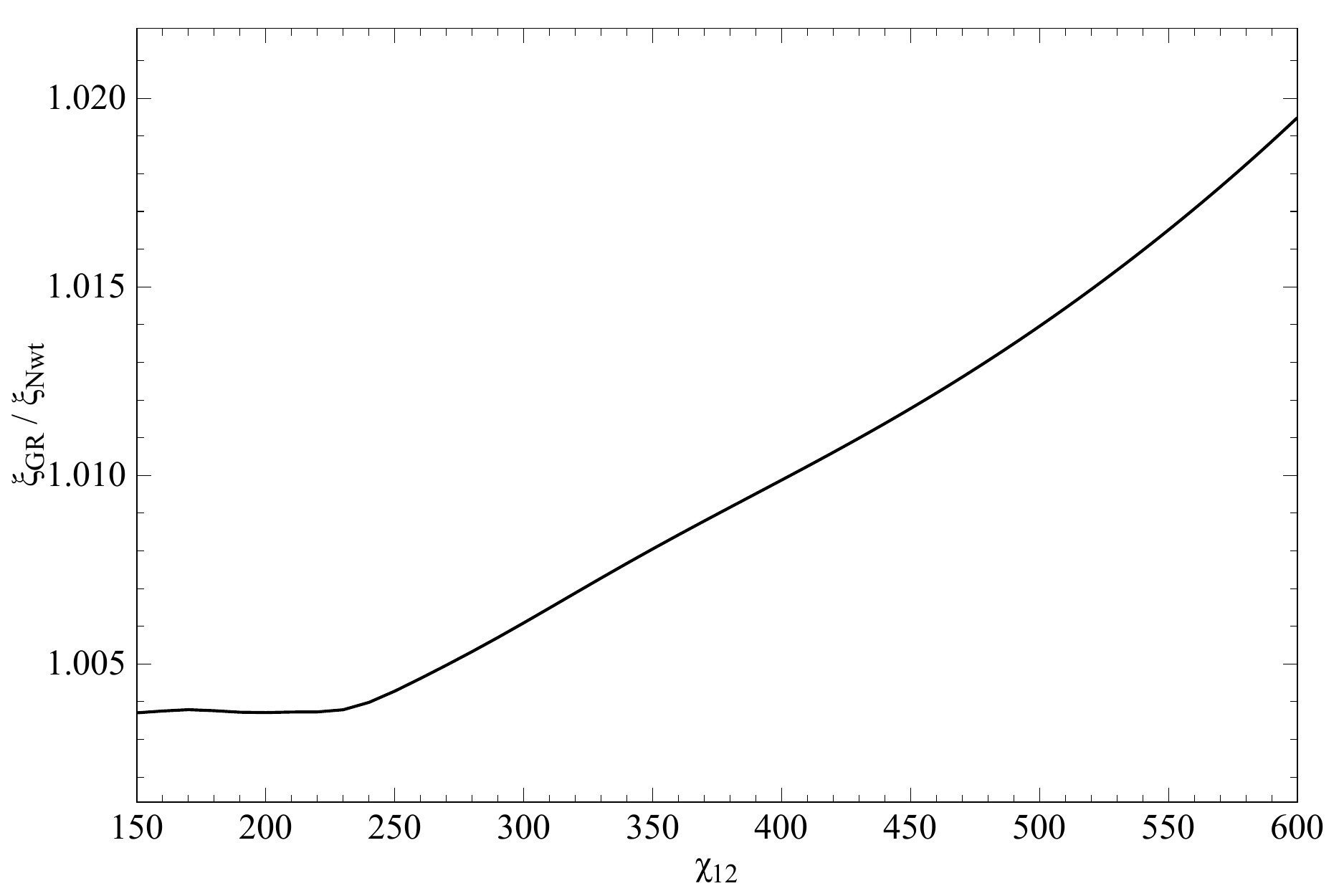}~~~~
\includegraphics[width=0.47\columnwidth]{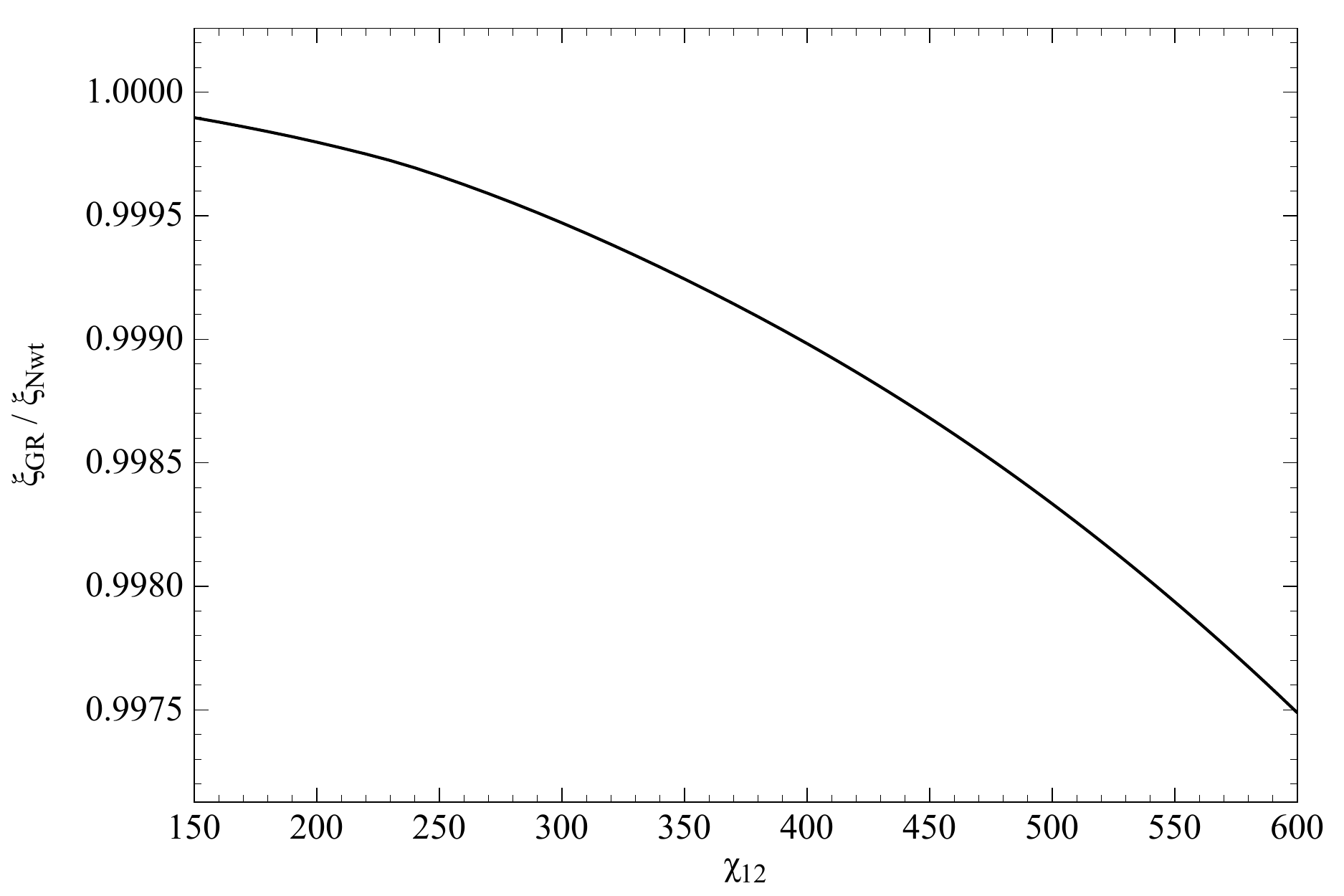}
\caption{The GR corrections to the Newtonian plane-parallel limit of the $\xi_{ss}$ monopole (left) and quadrupole (right), as a function of galaxy separation (Mpc).}\label{pplxi}
\end{figure*}
\end{center}

In Fig. \ref{pplxi}, we illustrate the GR corrections in \eqref{ppxiss} to the Newtonian monopole and quadrupole of $\xi_{ss}$. For galaxy separations up to $600\,$Mpc, the GR corrections to the quadrupole remain negligible while the monopole is corrected by up to 2\%.

Note that in the Newtonian limit, a monopole and a quadrupole term proportional to $\alpha^2$ survive:
\begin{eqnarray}
\frac{\beta_1^2 \alpha_1^2}{3\chi_1^2}\, \xi_0^{2}\, \mathcal{P}_{0}( { {{\bf n}}}_1 \cdot  { {{\bf n}}}_{11})~~~ \mbox{and} ~~~ - \frac{2\beta_1^2\alpha_1^2}{3\chi_1^2} \, \xi_2^{2}\, \mathcal{P}_{2}( { {{\bf n}}}_1 \cdot  { {{\bf n}}}_{11});
\end{eqnarray}
see also \cite{Szalay:1997cc}. These terms are usually omitted from the standard flat-sky analyses \cite{Kaiser:1987qv, Hamilton:1997zq} (for details see \cite{Raccanelli:2010hk}).

The remaining correlation functions $\xi_{AB}$, which have not before been explicitly computed in the plane-parallel limit, are presented in Appendix \ref{ppl}.



\subsection*{GR corrections on very large scales: analytical approximations}

For simplicity we consider the angular correlation, with $z_1=z_2\equiv z$.
We rewrite $\xi_L^{n}$ as
\begin{equation}\label{xiLn2}
\xi_L^{n}(\chi_{12}; z) = \int_{k_{\rm min}}^{1/\chi_{12}}  \frac{dk}{2\pi^2} k^{2-n}
j_L(\chi_{12} k) \, P_\delta(k; z)+  \int_{1/\chi_{12}}^\infty  \frac{dk}{2\pi^2} k^{2-n}
j_L(\chi_{12} k) \, P_\delta(k; z),
\end{equation}
where we impose a large-scale cutoff $k_{\rm min} $, which we take as $k_{\rm min} \sim H_0/2$. We take $\chi_{12} \sim \chi_{H}\simeq 2(1+z)H^{-1}(z)$.  In this case, for $k < 1/ \chi_{12}$, we have $P_\delta(k; z)\propto P_{\rm prim}(k) =A k^{n_s}$, and $j_L(\chi_{12} k) \simeq (\chi_{12} k)^L/ (2L+1)!!$. The second integral, for $L>0$, can be approximated  as
\begin{eqnarray}
&& \int_{1/\chi_{12}}^\infty  \frac{dk}{2\pi^2} k^{2-n} j_L(\chi_{12} k) \, P_\delta(k; z) \sim \frac{k_L^{2-n}}{2 \pi^2} \frac{P_\delta(k_L; z)}{\chi_{12}} I_L\;,\\
&& k_L= {(L+1/2)\over \chi_{12}},~~~
I_L = \int_0^\infty  j_L(y) dy = \frac{\sqrt{\pi}}{2}
\frac{\Gamma[(L+1)/2]}{\Gamma[(L+2)/2]}\;.
\end{eqnarray}
For $L=0$, the integral vanishes because $k_L<1/\chi_{12}$. Also, we can take $P_\delta \propto P_{\rm prim}(k) $ since $k_L \ll k_{\rm eq}$.
Then, for the spectral index $n_s < 1$, we obtain the analytical approximation:
\begin{eqnarray}
\label{xi_chi12}
\xi_L^{n}(\chi_{12}; z) \propto  \;
\left\{ \begin{array}{ll}
(3+n_s-n)^{-1} \left(\chi_{12}^{n-3-n_s}-\chi_{H_0}^{n-3-n_s}\right) & {\rm for}~~L=0 \\ \\
(3+L+n_s-n)^{-1}\left[ \left(\chi_{12}^{n-3-L-n_s}-\chi_{H_0}^{n-3-L-n_s}\right) + (L+1/2)^{2-n+ns} I_L \, \chi_{12}^{n-3-n_s}\right] & {\rm for}~~ L>0.
\end{array} \right. 
\end{eqnarray}
Note that, when $n=4$ and $L=0$, we require an infrared (IR) cutoff, $k_{\rm min} > 0$, since $\xi_0^4$ becomes power-law divergent. (If $n_s=1$, there is a logarithmic divergence.) 
The IR cutoff appears only in the terms of the correlation function
that contain $Y_{LM}$ with $M=L=0$. (In this case $Y_{00} \propto \mathcal{P}_0 \equiv 1$.) 
Therefore they add only an overall additive normalization to $\xi$ which is unobservable, and so the IR
cutoff actually conveys no information.
In addition, for large $\chi_{12}$, the slope of $\xi_L^{n}$ is less steep when $n$ increases. In other words, on large scales,  $B_{ss}$ are bigger when they contain GR corrections. (The same is true for the other $B_{AB}$.)

Defining $\theta$ as the angular separation of two galaxies,  we have $\chi_{12}=[2(1-\cos{\theta})]^{1/2}\chi$. We assume that $n_g \sim \chi_{12}^{-\nu}$, where $\nu$ is a suitable positive constant.
Then by \eqref{alpha} and \eqref{gamma}, in the large-scale limit $\chi_{12}\gg \chi_H(z)$, we have
\begin{eqnarray}
{\alpha \over \chi} & \to & - {2 \over \chi_H(z)}\left[2+\frac{3}{2}\Omega_m(z) \right],\\
\gamma &\to & -3\frac{\Omega_m(z)}{b(z)\chi_H^2(z)}\left[3\Omega_m(z)+8-2f(z)\right]\;.
\end{eqnarray}

\subsection*{Computation of GR wide-angle correlation function}

\begin{center}
\begin{figure*}[htb!]
\includegraphics[width=0.47\columnwidth]{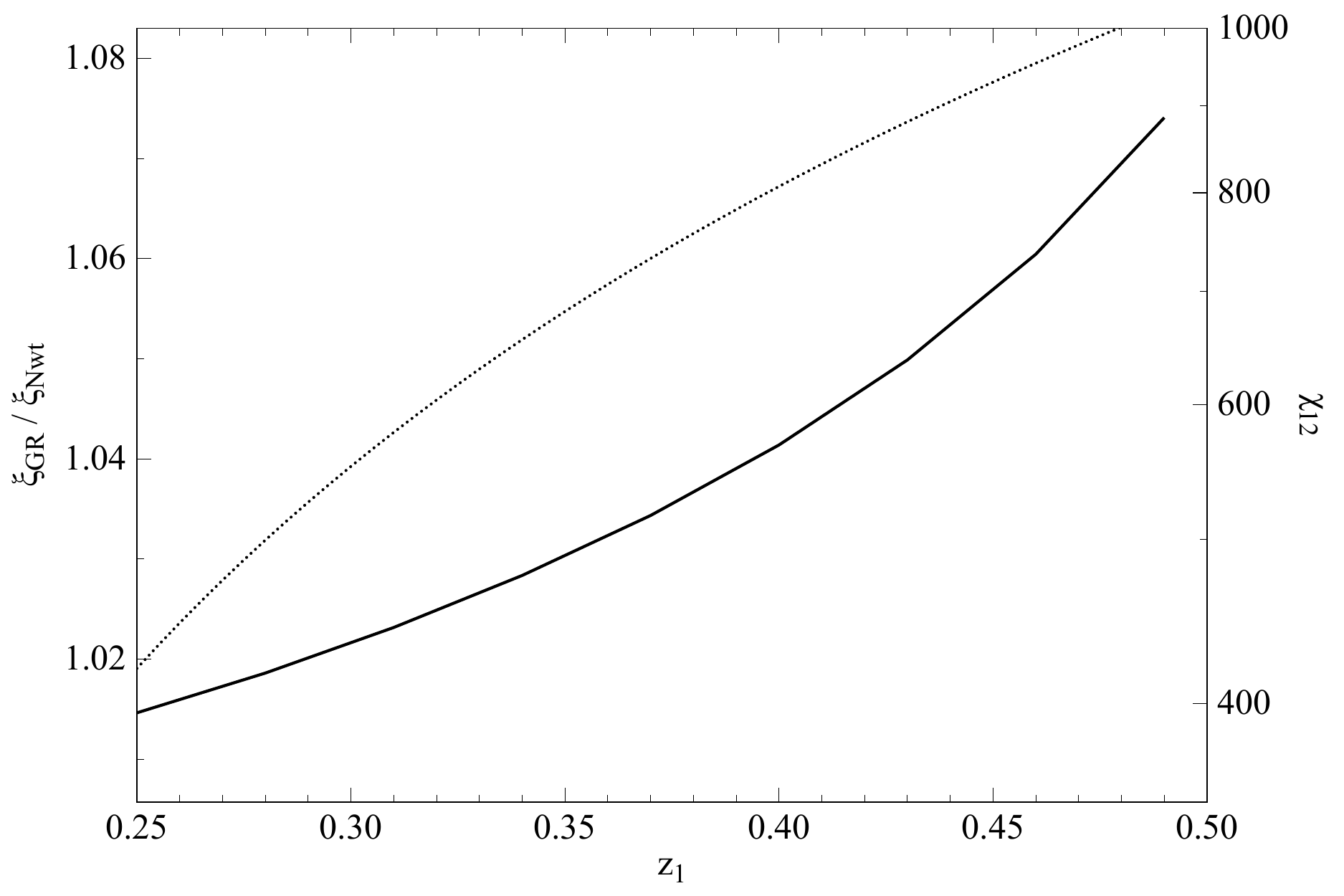}~~~~
\includegraphics[width=0.47\columnwidth]{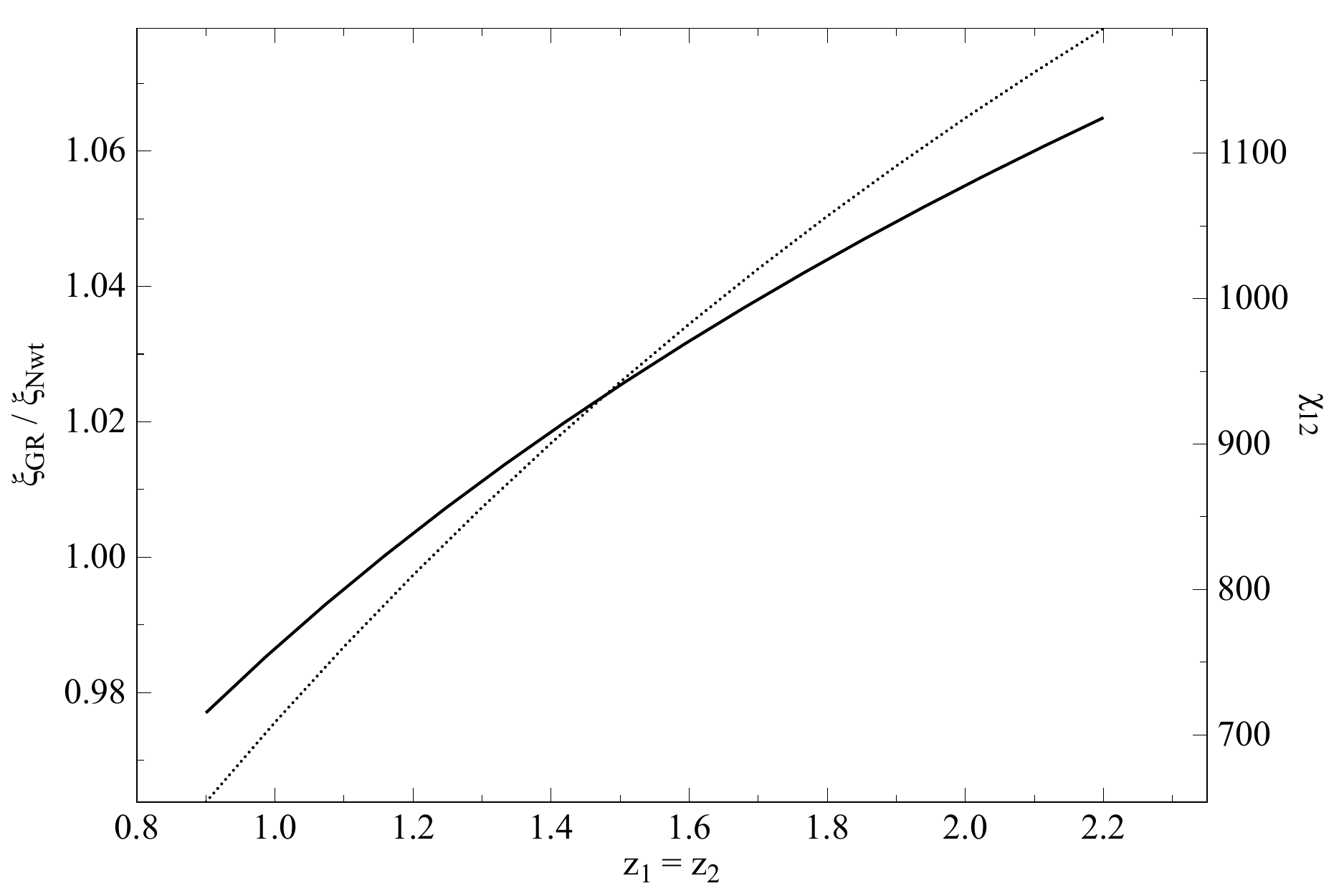}
\caption{
GR corrections to the Newtonian wide-angle $\xi_{ss}$ correlation function (solid lines), along the line of sight (left; with $z_2=0.1$ and $\theta = 0.1 \, \mbox{rad}$) and transverse (right; 
with $\theta = 0.3 \, \mbox{rad}$).
The corresponding comoving galaxy separation (Mpc) is also shown (dotted lines).
}
\label{waxi}
\end{figure*}
\end{center}

Finally, here we show the GR corrections to the full wide-angle correlation function. 
For illustrative purposes, we split the effects along the line of sight (${\bf n}\cdot\hat{\bf k}=1$), and transverse to the line of sight (${\bf n}\cdot\hat{\bf k}=0$).
The line of sight case corresponds to a configuration where one  galaxy is much further than the other and the angular separation $\theta$ is small. In the transverse case,  the two galaxies have the same redshift and $\theta$ is not small.
In Fig. \ref{waxi} we plot the GR corrections for the two examples:
\begin{itemize}
\item
$\xi_{ss}$, as a function of $z_1$, computed for $z_2=0.1$ and $\theta = 0.1 \, \mbox{rad}$.
\item
$\xi_{ss}$, as a function of $z_1 = z_2$, computed for $\theta = 0.3 \, \mbox{rad}$.
\end{itemize}
Along with the correlation function corrections, we plot the comoving galaxy separation, in order to have a clearer picture of the scales involved.
In these computations we assumed a constant bias $b= 2$ and $n_g$ from the universal mass function prescription, see \cite{Jeong:2011as}.
The plots of the GR corrections show the ratio of the correlation function computed using the GR formalism with the one in the Newtonian case.
Wide-angle effects are included in both cases, the only difference being the inclusion or not of terms deriving from $\gamma$.
We decided to analyze two different cases, both for uniformity with previous literature and for physical reasons: pairs of galaxies (almost) along, and across the line of sight.
The first case (left panel) corresponds to pairs that have a large difference in redshift and experience the maximum of the redshift-space distortions effect, while the second one (right panel) corresponds to pairs of galaxies with the same redshift, and for which the redshift space distortion effect is minimum.
In a following companion paper we will present a detailed analysis of all the GR effects and their dependencies on scale, angular separation, bias and radial distribution of sources for all the terms presented in this work, along with predictions for observing these effects with future surveys.
%


 \section{Conclusions}
\label{conclusions}

 We derived a fully general relativistic expression for correlation functions in the wide-angle case,  that will need to be used in future surveys (such as SKA and Euclid) which measure galaxy correlations on very large scales. Our formalism recovers and generalizes previous results in the plane-parallel (flat-sky) and Newtonian approximations. 
 
We presented new results for the GR-corrected plane-parallel case. The GR corrections to the Newtonian monopole of $\xi_{ss}$, for galaxy separations up to $600\,$Mpc, are up to 2\% (see Fig. \ref{pplxi}). 

Stronger GR corrections arise in the wide-angle case.
We showed via illustrative examples that GR corrections on large scales ($\sim 1\,$Gpc separation) can in principle be of order 5--10\%, as shown in Fig. \ref{waxi}. Of course, the observability of these effects is severely degraded by cosmic variance. Future large-volume surveys will reduce the cosmic variance -- but the problem can in fact be removed if we are able to observe multiple tracers of the underlying matter distribution, as shown in \cite{Seljak:2008xr}. In a companion paper we will present a detailed analysis of the predicted GR corrections for large-volume surveys, taking into account cosmic variance.

\vspace{0.5in}

{\bf Acknowledgments:}\\
DB and RM are supported by the South African Square Kilometre
Array Project. RM is supported by the STFC (UK) (grant no.
ST/H002774/1). RM and CC are supported by the National Research Foundation
(NRF, South Africa). DB, RM and CC are supported by a Royal
Society (UK)/ NRF (SA) exchange grant. Part of the research
described in this paper was carried out at the Jet Propulsion
Laboratory, California Institute of Technology, under a contract
with the National Aeronautics and Space Administration.

\appendix

\section{Coefficients in the tripolar decomposition of $\xi_{AB}$}
\label{bco}

For (\ref{eq:xi_ss}):
\begin{eqnarray}
\begin{array} {lll}
 B_{ss \; 0}^{\phantom{ss \;}000} =\left(1+\frac{1}{3}\beta_1\right) \left(1+\frac{1}{3} \beta_2\right),  & \quad&
 B_{ss \; 1}^{\phantom{ss \;}011} = \sqrt{3}\; \chi_2^{-1} \left(1+\frac{1}{3}\beta_1\right)  \beta_2\alpha_2 ,  \\ \\
 B_{ss \; 2}^{\phantom{ss \;}000} =- \left(1+\frac{1}{3}\beta_1\right) \gamma_2 - \left(1+\frac{1}{3} \beta_2\right)\gamma_1, & \quad &
 B_{ss \; 0}^{\phantom{ss \;}022} =- \frac{2 \sqrt{5}}{3} \left(1+\frac{1}{3}\beta_1\right) \beta_2 ,  \\ \\
 B_{ss \; 1}^{\phantom{ss \;}101} = -  \sqrt{3}  \;\chi_1^{-1} \beta_1\alpha_1  \left(1+\frac{1}{3} \beta_2\right), & \quad&
 B_{ss \; 2}^{\phantom{ss \;}110} =  - \frac{\sqrt{3}}{3} \;
{\chi}_1^{-1} {\chi}_2^{-1} \beta_1\alpha_1 \beta_2\alpha_2  ,   \\ \\ 
B_{ss \; 2}^{\phantom{ss \;}112} =  - \frac{\sqrt{30}}{3} \; {\chi}_1^{-1} {\chi}_2^{-1} \beta_1\alpha_1 \beta_2\alpha_2,  & \quad&
 B_{ss \; 3}^{\phantom{ss \;}101} =  \sqrt{3}  \; {\chi}_1^{-1} \beta_1\alpha_1\gamma_2 ,  \\ \\
B_{ss \; 1}^{\phantom{ss \;}121} =  \frac{2\sqrt{30}}{15} \; {\chi}_1^{-1} \alpha_1\beta_1\beta_2 ,  & \quad&
 B_{ss \;1}^{\phantom{ss \;}123} =   \frac{2\sqrt{105}}{15} \; {\chi}_1^{-1} \alpha_1\beta_1 \beta_2,  \\ \\
B_{ss \; 3}^{\phantom{ss \;}011} = -  \sqrt{3} \;  \chi_2^{-1} \gamma_1 \beta_2 \alpha_2, & \quad&
 B_{ss \; 4}^{\phantom{ss\;}000} =   \gamma_1   \gamma_2 , \\ \\
 B_{ss \; 2}^{\phantom{ss \;}022} =   \frac{2 \sqrt{5}}{3}  \; \gamma_1 \beta_2  ,  & \quad&
B_{ss \; 1}^{\phantom{ss \;}211} = -\frac{2 \sqrt{30}}{15}   \; \chi_2^{-1} \beta_1 \beta_2 \alpha_2  ,  \\ \\
B_{ss \; 1}^{\phantom{ss \;}213} = - \frac{2\sqrt{105}}{15}  \; \chi_2^{-1} \beta_1\beta_2 \alpha_2 ,  & \quad&
 B_{ss \;0}^{\phantom{ss \;}202} =    - \frac{2 \sqrt{5}}{3}  \; \beta_1\left(1+\frac{1}{3}\beta_2\right) , \\ \\
 B_{ss \; 2}^{\phantom{ss \;}202} =\frac{2 \sqrt{5}}{3}\; \beta_1\gamma_2 , & \quad&
 B_{ss \; 0}^{\phantom{ss \;}220}=  \frac{4 \sqrt{5}}{45} \;  \beta_1  \beta_2 , \\ \\
 B_{ss \; 0}^{\phantom{ss \;}222} =  \frac{4 \sqrt{70}}{63}\beta_1  \beta_2 , & \quad&
 B_{ss \; 0}^{\phantom{ss \;}224} = \frac{4 \sqrt{70}}{35}   \beta_1  \beta_2 \;,  \\
  \end{array}  \nonumber \\
\end{eqnarray}
where $\alpha_i=\alpha(z_i)$, $\beta_i=\beta(z_i)$ and $\gamma_i=\gamma(z_i)$.

For (\ref{eq:xi_kk}):
\begin{eqnarray} 
\begin{array} {lll} 
B_{\kappa \kappa \; 0}^{\phantom{\kappa \kappa \;}000}  = \sigma_{1\tilde 1} \sigma_{2\tilde 2}, & 
\quad & 
B_{\kappa \kappa \; 0}^{\phantom{\kappa \kappa \;}022} =\sqrt{5} \; \sigma_{1\tilde 
1}\sigma_{2\tilde 2},   \\ \\ 
B_{\kappa \kappa \; 0}^{\phantom{\kappa \kappa \;}202}= \sqrt{5} \; \sigma_{1\tilde 1} 
\sigma_{2\tilde 2}, & \quad & 
B_{\kappa \kappa \; 1}^{\phantom{\kappa \kappa \;}011}  =-3 \sqrt{3} \; \tilde\chi_2^{-1} 
\;\sigma_{1\tilde 1} \sigma_{2\tilde 2} ,   \\ \\ 
B_{\kappa \kappa \; 1}^{\phantom{\kappa \kappa \;}101} =3\sqrt{3}  \; \tilde\chi_1^{-1} \; 
\sigma_{1\tilde 1}\sigma_{2\tilde 2} , & \quad & 
B_{\kappa \kappa \;0}^{\phantom{\kappa \kappa \;}220} = \frac{\sqrt{5} }{ 5}\; \sigma_{1\tilde 1} 
\sigma_{2\tilde 2} , \\ \\ 
B_{\kappa \kappa \; 0}^{\phantom{\kappa \kappa \;}222} =\frac{\sqrt{70} }{ 7} \; \sigma_{1\tilde 1} 
\sigma_{2\tilde 2} , & \quad & 
B_{\kappa \kappa \; 0}^{\phantom{\kappa \kappa \;}224} =\frac{9\sqrt{70} }{ 35} \; \sigma_{1\tilde 1} 
\sigma_{2\tilde 2} ,   \\ \\ 
B_{\kappa \kappa \; 1}^{\phantom{\kappa \kappa \;}211} =-\frac{3 \sqrt{30} }{5} \; \tilde\chi_2^{-1} 
\; \sigma_{1\tilde 1} \sigma_{2\tilde 2} , & \quad & 
B_{\kappa \kappa \; 1}^{\phantom{\kappa \kappa \;}213}=- \frac{3\sqrt{105} }{5} \; \tilde\chi_2^{-1} 
\;  \sigma_{1\tilde 1} \sigma_{2\tilde 2},   \\ \\ 
B_{\kappa \kappa \; 1}^{\phantom{\kappa \kappa \;}121}= \frac{3\sqrt{30} }{5} \; \tilde\chi_1^{-1} \; 
\sigma_{1\tilde 1}\sigma_{2\tilde 2} , & \quad & 
B_{\kappa \kappa \;1}^{\phantom{\kappa \kappa \;}123}  = \frac{3 \sqrt{105}}{5}\; \tilde\chi_1^{-1} \; 
\sigma_{1\tilde 1} \sigma_{2\tilde 2} ,   \\ \\ 
B_{\kappa \kappa \; 2}^{\phantom{\kappa \kappa \;}110} =-3\sqrt{3}\; \tilde\chi_1^{-1} 
\tilde\chi_2^{-1}\; \sigma_{1\tilde 1} \sigma_{2\tilde 2}  , & \quad & 
B_{\kappa \kappa \; 2}^{\phantom{\kappa \kappa \;}112}= - 3\sqrt{30} \; \tilde\chi_1^{-1} 
\tilde\chi_2^{-1}\; \sigma_{1\tilde 1} \sigma_{2\tilde 2}  , \\ \\ 
\end{array} \nonumber \\ \label{Bkk} 
\end{eqnarray} 
where $\sigma_{i\,\tilde i} = \sigma( {z}_i, \tilde z_i)$.

For (\ref{eq:xi_II}):
\begin{eqnarray}
B_{II \; 4}^{\phantom{II \;}000}& =& \mu_{1\tilde 1}  \mu_{2\tilde 2}, \label{BII}
\end{eqnarray}
where $ \mu_{i\,\tilde i} = \mu( {z}_i,\tilde z_i)$.

For (\ref{eq:xi_sI}):
\begin{eqnarray}
\begin{array} {lll}
B_{sI \; 2}^{\phantom{sI \;}000} = -\left(1+\frac{1}{3}\beta_1\right) \mu_{2\tilde 2} , & \quad &
B_{sI \;3}^{\phantom{sI \;}101} = \sqrt{3} \;  \chi_1^{-1} \beta_1 \alpha_1 \mu_{2\tilde 2},    \\ \\
B_{sI \; 4}^{\phantom{sI \;}000}= \gamma_1 \mu_{2\tilde 2} , & \quad &
 B_{sI \; 2}^{\phantom{sI \;}202} =\frac{2 \sqrt{5}}{3} \; \beta_1 \mu_{2\tilde 2} .
\end{array} \nonumber \\
\end{eqnarray}

For (\ref{eq:xi_sk}):
\begin{eqnarray} 
\begin{array} {lll} 
B_{s\kappa \; 0}^{\phantom{s\kappa \;}000}= \left(1+\frac{1}{3}\beta_1\right)\sigma_{2\tilde 2}  , & 
\quad & 
B_{s\kappa \;0}^{\phantom{s\kappa \;}022} = \sqrt{5} \;\left(1+\frac{1} 
{3}\beta_1\right)\sigma_{2\tilde 2} ,    \\ \\ 
B_{s\kappa \; 1}^{\phantom{s\kappa \;}011}= - \; 3\sqrt{3}\,\tilde\chi_2^{-1} \left(1+\frac{1} 
{3}\beta_1\right)\sigma_{2\tilde 2}  , & \quad & 
B_{s\kappa \;1}^{\phantom{s\kappa \;}101} = -  \sqrt{3} \;\chi_1^{-1}\, \beta_1 \alpha_1 
\sigma_{2\tilde 2}  ,    \\ \\ 
B_{s\kappa \; 1}^{\phantom{s\kappa \;}121} = -\frac{\sqrt{30}}{5} \; \chi_1^{-1}\, \beta_1 \alpha_1 
\sigma_{2\tilde 2} , & \quad & 
B_{s\kappa \;1}^{\phantom{s\kappa \;}123} = - \frac{\sqrt{105}}{5} \;  \chi_1^{-1}\, \beta_1 \alpha_1 
\sigma_{2\tilde 2}  , \\ \\ 
B_{s\kappa \; 2}^{\phantom{s\kappa \;}110} = \sqrt{3}\, \tilde\chi_2^{-1} \chi_1^{- 
1}\,\beta_1\alpha_1\sigma_{2\tilde 2} , & \quad & 
B_{s\kappa \;2}^{\phantom{s\kappa \;}112} = \sqrt{30}\; \tilde\chi_2^{-1} \chi_1^{- 
1}\,\beta_1\alpha_1  \sigma_{2\tilde 2} ,\\ \\ 
B_{s\kappa \; 2}^{\phantom{s\kappa \;}000} = - \; \gamma_1 \;  \sigma_{2\tilde 2}  , & \quad & 
B_{s\kappa \;2}^{\phantom{s\kappa \;}022} = - \sqrt{5} \;  \gamma_1 \; \sigma_{2\tilde 2}  ,    \\ \\ 
B_{s\kappa \; 3}^{\phantom{s\kappa \;}011} =   3\sqrt{3} \tilde\chi_2^{-1}     \gamma_1 
\sigma_{2\tilde 2},  & \quad & 
B_{s\kappa \; 0}^{\phantom{s\kappa \;}202} =-\frac{2\sqrt{5}}{3}\;  \beta_1\sigma_{2\tilde 2} ,    \\ 
\\ 
B_{s\kappa \; 0}^{\phantom{s\kappa \;}220}=-\frac{2\sqrt{5}}{15}\;  \beta_1 \sigma_{2\tilde 2}  , & 
\quad & 
B_{s\kappa \; 0}^{\phantom{s\kappa \;}222} =-\frac{2\sqrt{70}}{21} \; \beta_1\sigma_{2\tilde 2} ,    \\ 
\\ 
B_{s\kappa \; 0}^{\phantom{s\kappa \;}224}=-\frac{6\sqrt{70}}{35}  \;  \beta_1 \sigma_{2\tilde 2}, & 
\quad & 
B_{s\kappa \; 1}^{\phantom{s\kappa \;}211} =\frac{2\sqrt{30}}{5} \tilde\chi_2^{-1}\beta_1 
\sigma_{2\tilde 2} , \\ \\ 
B_{s\kappa \; 1}^{\phantom{s\kappa \;}213} =\frac{2\sqrt{105}}{5} \tilde\chi_2^{-1}\;   \beta_1   
\sigma_{2\tilde 2} . 
\end{array} \nonumber \\ 
\end{eqnarray} 

For (\ref{eq:xi_kI}):
\begin{eqnarray}
\begin{array} {lll}
B_{\kappa I \; 2}^{\phantom{\kappa I  \;}000} =-\; \sigma_{1\tilde 1}\mu_{2\tilde 2} , & \quad &
 B_{\kappa I \; 2}^{\phantom{\kappa I  \;}202} =-\sqrt{5} \; \sigma_{1\tilde 1}\mu_{2\tilde 2},   \\ \\
B_{\kappa I \; 3}^{\phantom{\kappa I  \;}101} = - 3\sqrt{3} \; \tilde \chi_1^{-1}  \; \sigma_{1\tilde 1} \mu_{2\tilde 2} .
\end{array} \nonumber \\
\end{eqnarray}

The remaining $\xi_{AB}$ are
\begin{eqnarray}
{\xi}_{Is}( {\bf x}_1, {\bf x}_2) &=& b(z_1)b(z_2)
\int^{ {\chi}_1}{d\tilde\chi_1} \sum_{\ell_1,\ell_2,L,n} B_{Is \;
n}^{\phantom{sI \;}\ell_1\ell_2L}( {\chi}_1,\tilde {\chi}_1;
\chi_2) S_{\ell_1\ell_2L}( { {{\bf n}}}_1, { {{\bf n}}}_2, {{\bf
n}}_{ \tilde{1}2}) \, \xi_L^{n}(\chi_{ \tilde{1}2}; \tilde{z}_1,
{z}_2) , \label{eq:xi_Is}\\
{\xi}_{\kappa s}( {\bf x}_1, {\bf x}_2) &=&
b(z_1)b(z_2) \int^{ {\chi}_1}{d\tilde\chi_1}
\sum_{\ell_1,\ell_2,L,n} B_{\kappa s \; n}^{\phantom{sI
\;}\ell_1\ell_2L}( {\chi}_1,\tilde {\chi}_1; \chi_2)
S_{\ell_1\ell_2L}( { {{\bf n}}}_1, { {{\bf n}}}_2, {{\bf n}}_{
\tilde{1}2}) \, \xi_L^{n}(\chi_{ \tilde{1}2}; \tilde{z}_1, {z}_2)
, \label{eq:xi_ks}\\
{\xi}_{I\kappa}( {\bf x}_1, {\bf x}_2) &=&
b(z_1)b(z_2)
\int^{ {\chi}_1,\chi_2} {d\tilde\chi_1}{d\tilde\chi_2}
\sum_{\ell_1,\ell_2,L,n} B_{I\kappa \; n}^{\phantom{\kappa \kappa
\;}\ell_1\ell_2L}( {\chi}_1,\tilde\chi_1; {\chi}_2,
\tilde\chi_2)\, S_{\ell_1\ell_2L}( { {{\bf n}}}_1, { {{\bf n}}}_2,
{\tilde{\bf n}}_{12})\, \xi_L^{n}(\tilde\chi_{12}; \tilde{z}_1,
\tilde{z}_2)\!, \label{eq:xi_Ik}
\end{eqnarray}
where
\begin{eqnarray}
\chi_{ \tilde{1}2}{{\bf n}}_{ \tilde{1}2}&=& {\left(
\tilde{\chi}_1-\chi_1 \right) { {{\bf n}}}_1 +  {\chi}_{12} {
{{\bf n}}}_{12}},\\
\chi_{\tilde{1}2}^2 &=&{  \tilde{\chi}_1^2+ \chi_2^2 + \frac{
\tilde\chi_1}{ {\chi}_1}\left[ {\chi}_{12}^2-\left( {\chi}_1^2+
{\chi}_2^2\right)\right]},
\end{eqnarray}
and the corresponding $B$ coefficients are
\begin{eqnarray}
\begin{array} {lll}
B_{Is \; 2}^{\phantom{sI \;}000} = -\left(1+\frac{1}{3}\beta_2\right) \mu_{1\tilde 1} , & \quad &
 B_{Is \;3}^{\phantom{sI \;}011} = -\sqrt{3} \; \beta_2 \alpha_2 \chi_2^{-1}\mu_{1\tilde 1} ,    \\ \\
B_{Is \; 4}^{\phantom{sI \;}000} = \gamma_2 \, \mu_{1\tilde 1} , & \quad &
B_{Is \; 2}^{\phantom{sI \;}022} =\frac{2 \sqrt{5}}{3} \; \beta_2 \mu_{1\tilde 1} ,
\end{array} \nonumber \\
 \end{eqnarray}
\begin{eqnarray} 
\begin{array} {lll} 
B_{\kappa s \; 0}^{\phantom{s\kappa \;}000} =\left(1+\frac{1}{3}\beta_2\right) \sigma_{1\tilde 1}  ,  & 
\quad & 
B_{\kappa s \; 0}^{\phantom{s\kappa \;}202} =\sqrt{5} \; \left(1+\frac{1} 
{3}\beta_2\right)\sigma_{1\tilde 1} ,    \\ \\ 
B_{\kappa s \; 1}^{\phantom{s\kappa \;}101} =  3\sqrt{3} \; \tilde\chi_1^{-1}\left(1+\frac{1} 
{3}\beta_2\right)\sigma_{1\tilde 1}  ,  & \quad & 
B_{\kappa s \;1}^{\phantom{s\kappa \;}011} =   \sqrt{3} \;\chi_2^{-1}\beta_2 \alpha_2 \sigma_{1\tilde 
1}  ,    \\ \\ 
B_{\kappa s \; 1}^{\phantom{s\kappa \;}211}= \frac{\sqrt{30}}{5}\;  \chi_2^{-1} \beta_2\alpha_2 
\sigma_{1\tilde 1} , & \quad & 
B_{\kappa s \; 1}^{\phantom{s\kappa\;}213} =  \frac{\sqrt{105}}{5} \;  \chi_2^{-1}\, \beta_2\alpha_2 
\sigma_{1\tilde 1}  , \\ \\ 
B_{\kappa s \; 2}^{\phantom{s\kappa \;}110}= \sqrt{3}\; \tilde\chi_1^{-1}   \chi_2^{-1}\, 
\beta_2\alpha_2 \sigma_{1\tilde 1} , & \quad & 
B_{\kappa s \;2}^{\phantom{s\kappa \;}112} = \sqrt{30} \; \tilde\chi_1^{-1}  \chi_2^{-1} 
\beta_2\alpha_2  \sigma_{1\tilde 1} , \\ \\ 
B_{\kappa s \; 2}^{\phantom{s\kappa \;}000} = - \; \gamma_2 \;  \sigma_{1\tilde 1}  , & \quad & 
B_{\kappa s \;2}^{\phantom{s\kappa \;}202} = - \sqrt{5} \;  \gamma_2 \;\sigma_{1\tilde 1}  ,   \\ \\ 
B_{\kappa s \; 3}^{\phantom{s\kappa \;}101} =   - 3\sqrt{3}\tilde\chi_1^{-1}  \;   \gamma_2 
\sigma_{1\tilde 1}, & \quad & 
B_{\kappa s \; 0}^{\phantom{s\kappa \;}022} = -\frac{2\sqrt{5}}{3}\;  \beta_2\sigma_{1\tilde 1}  ,    \\ 
\\ 
B_{\kappa s \; 0}^{\phantom{s\kappa \;}220} =-\frac{2\sqrt{5}}{15}\;  \beta_2 \sigma_{1\tilde 1}  , & 
\quad & 
B_{\kappa s \; 0}^{\phantom{s\kappa \;}222} =-\frac{2\sqrt{70}}{21} \; \beta_2\sigma_{1\tilde 1} ,     
\\ \\ 
B_{\kappa s \; 0}^{\phantom{s\kappa \;}224} =-\frac{6\sqrt{70}}{35}  \;  \beta_2 \sigma_{1\tilde 1},  & 
\quad & 
B_{\kappa s \; 1}^{\phantom{s\kappa \;}121} =-\frac{2\sqrt{30}}{5 } \tilde\chi_1^{-1}\; \beta_2 
\sigma_{1\tilde 1},\\ \\ 
B_{\kappa s \; 1}^{\phantom{s\kappa \;}123}=-\frac{2\sqrt{105}}5 \tilde\chi_1^{-1}\;   \beta_2   
\sigma_{1\tilde 1} \;. 
\end{array} \nonumber \\ 
\end{eqnarray} 
\begin{eqnarray}
\begin{array} {lll}
B_{I\kappa  \; 2}^{\phantom{\kappa I  \;}000} =-\; \mu_{1\tilde 1}\sigma_{2\tilde 2} , & \quad &
 B_{I\kappa  \; 2}^{\phantom{\kappa I  \;}022} =-\sqrt{5} \; \mu_{1\tilde 1}\sigma_{2\tilde 2},   \\ \\
B_{I\kappa  \; 3}^{\phantom{\kappa I  \;}011} =  3\sqrt{3} \; \tilde\chi_2^{-1} \mu_{1\tilde 1}\sigma_{2\tilde 2} .
\end{array} \nonumber \\
\end{eqnarray}

\section{General relativistic $\xi_{AB}$ in the plane-parallel limit}
\label{ppl}

$\xi_{ss}$ is given in \eqref{ppxiss}. The remaining GR $\xi_{AB}$ have not before been given in the plane-parallel limit.

Lensing-lensing correlation:
\begin{eqnarray} 
{\xi}_{\kappa \kappa}( {\bf x}_1, {\bf x}_1) &=& b_1^2 \int^{ {\chi}_{1}} {d\chi_{1'}}\int^{ 
{\chi}_1}{d\chi_{1''}} \sigma_{11'} \sigma_{11''}  \bigg\{  \nonumber \\ 
&& \left[ 3\left(\frac{2}{5}\xi_0^{0}( \chi_{11'}; z_{1'}, z_{1''}) + \frac{1}{\chi_{1'} \chi_{1''}} \xi_0^{2}( 
\chi_{1'1''}; z_{1'}, z_{1''})\right) \mathcal{P}_{0}( { {{\bf n}}}_1 \cdot  {\bf n}_{1'1''})  \right. \nonumber \\ 
&+& 3 \frac{\chi_{1''}-\chi_{1'}}{\chi_{1'} \chi_{1''}} \xi_1^{1}( \chi_{1'1''}; z_{1'}, z_{1''}) \mathcal{P}_{1}( { 
{{\bf n}}}_1 \cdot {\bf n}_{1'1''})  \nonumber \\ 
&+&6 \left(\frac{2}{7}\xi_2^{0}( \chi_{1'1''}; z_{1'}, z_{1''}) - \frac{1}{\chi_{1'} \chi_{1''}} \xi_2^{2}( 
\chi_{1'1''}; z_{1'}, z_{1''})\right) \mathcal{P}_{2}( { {{\bf n}}}_1 \cdot  {\bf n}_{1'1''})   \nonumber \\ 
&-& \left. \frac{9}{5}\frac{\chi_{1''}-\chi_{1'}}{\chi_{1'}\chi_{1''}} \xi_3^{1}( \chi_{1'1''}; z_{1'}, 
z_{1''})\mathcal{P}_{3}( { {{\bf n}}}_1 \cdot  {\bf n}_{1'1''})+\frac{18}{35} \xi_4^{0}( \chi_{1'1''}; 
z_{1'},z_{1''})\mathcal{P}_{4}( { {{\bf n}}}_1 \cdot  {\bf n}_{1'1''}) 
\right] \bigg\}\;. 
\end{eqnarray}

$II$ correlation:
\begin{eqnarray}
 {\xi}_{II}( {\bf x}_1, {\bf x}_1) = b_1^2 \int^{ {\chi}_1} {d\chi_{1'}}\int^{ {\chi}_1}{d\chi_{1''}} \mu_{11'} \mu_{11''} \; \xi_0^{4}( \chi_{1'1''}; z_{1'}, z_{1''})\mathcal{P}_{0}( { {{\bf n}}}_1 \cdot  {\bf n}_{1'1''}).
\end{eqnarray}
We assumed $ {\bf n}_{12} \equiv {\bf
n}_{1'1''}$. Then
\begin{eqnarray}
\chi_{1'1''} {\bf n}_{1'1''} &=& \left(\chi_{1'} - \chi_{1''} \right) {\bf n}_{1'} +\chi_{1''} {\chi}_1^{-1}  {\chi}_{11} {\bf n}_{11}\;,\\
\chi_{1'1''}^2 &=& \chi_{1'}^2+ \chi_{1''}^2 +\chi_{1'}\chi_{1''}\left(\frac{ {\chi}_{11}^2}{{\chi}_1^2}-2\right)\;.
\end{eqnarray}

${sI}$ correlation:
\begin{eqnarray}
{\xi}_{sI}( {\bf x}_1, {\bf x}_1)&=& b_1^2 \int^{ {\chi}_{1}} {d\chi_{1'}}  \mu_{11'} \bigg\{ \left[ - \left(1+\frac{1}{3}\beta_1 \right) \xi_0^{2}( \chi_{ {1} 1'};  {z}_{1}, z_{1'})
 +  \gamma_1 \xi_0^{4}( \chi_{ {1} 1'};  {z}_{1}, z_{1'})  \right] \mathcal{P}_{0}( { {{\bf n}}}_1 \cdot  {{\bf n}}_{ {1}1'}) \nonumber \\
 &&~- \frac{\beta_1 \alpha_1}{\chi_1} \xi_1^{3}  ( \chi_{ {1} 1'};  {z}_{1}, z_{1'}) \mathcal{P}_{1}( { {{\bf n}}}_1 \cdot  {{\bf n}}_{ {1}1'})  + \frac{2}{3} \beta_1  \xi_2^{2}  ( \chi_{ {1} 1'}; {z}_{1}, z_{1'}) \mathcal{P}_{2}( { {{\bf n}}}_1 \cdot  {{\bf n}}_{ {1}1'}) \bigg\}\;,
  \end{eqnarray}
where we assumed $ {\bf n}_{1\tilde{2}}\equiv {\bf
n}_{ {1}1'}$. Then
\begin{eqnarray}
\chi_{ {1}1'} {{\bf n}}_{ {1}1'} &=& \left(\chi_1 - \chi_{1'} \right) {\bf n}_1 +\chi_{1'} {\chi}_1^{-1}{\chi}_{11} {\bf n}_{11}\;,\\
\chi_{ {1}1'}^2 &=&  {\chi}_1^2+ \chi_{1'}^2 +  \frac{\chi_{1'}}{{\chi}_1}\left( {\chi}_{11}^2-2 {\chi}_1^2\right)\;.
\end{eqnarray}

${Is}$ correlation:
\begin{eqnarray}
{\xi}_{Is}( {\bf x}_1, {\bf x}_1)&=& b_1^2 \int^{ {\chi}_{1}} {d\chi_{1}}  \mu_{11'} \bigg\{ \left[ - \left(1+\frac{1}{3}\beta_1 \right) \xi_0^{2}( \chi_{1'1};  {z}_{1'}, z_{1}) +  \gamma_1 \xi_0^{4}( \chi_{1'1};  {z}_{1'}, z_{1})  \right] \mathcal{P}_{0}( { {{\bf n}}}_1 \cdot  {{\bf n}}_{1'1}) \nonumber \\
 &&~+ \frac{\beta_1 \alpha_1}{\chi_1}\xi_1^{3}  ( \chi_{1'1}; {z}_{1'}, z_{1}) \mathcal{P}_{1}( { {{\bf n}}}_1 \cdot  {{\bf n}}_{1' 1}) + \frac{2}{3} \beta_1  \xi_2^{2}  ( \chi_{1'1}; {z}_{1'}, z_{1}) \mathcal{P}_{2}( { {{\bf n}}}_1 \cdot  {{\bf n}}_{1'1}) \bigg\}\;,
 \end{eqnarray}
where we defined $ {\bf n}_{\tilde{1} 2} \equiv{\bf n}_{1'1}$. Then
\begin{equation}
\chi_{1'1} {\bf n}_{1'1}=-\left( {\chi}_1 - \chi_{1'} \right) {\bf n}_1 + {\chi}_{11} {\bf n}_{11}~~~\mbox{and}~~ \chi_{1' 1}=\chi_{ {1}1'}\;.
\end{equation}

${s\kappa}$ correlation:
\begin{eqnarray} 
{\xi}_{s\kappa}( {\bf x}_1, {\bf x}_1)&=& b_1^2 \int^{ {\chi}_{1}} {d\chi_{1'}}   \sigma_{11'} 
\bigg\{\bigg[ \left(1+\frac{1}{5}\beta_1 \right) \xi_0^{0}( \chi_{ {1} 1'};  {z}_{1}, z_{1'}) 
 - \left(\frac{\beta_1 \alpha_1}{\chi_{1'}{\chi}_1} +\gamma_1 \right) \xi_0^{2}( \chi_{ {1} 1'}; {z}_{1}, 
z_{1'}) \bigg] \mathcal{P}_{0}( { {{\bf n}}}_1 \cdot  {{\bf n}}_{ {1}1'}) \nonumber \\ 
&+&  \bigg[ 3 \left(\frac{1}{\chi_{1'}}+\frac{3}{5\,\chi_{1'}}\beta_1 + \frac{1}{5}\frac{\beta_1\alpha_1} 
{\chi_1}\right) \xi_1^{1}( \chi_{ {1} 1'};  {z}_{1}, z_{1'}) 
- \frac{3}{\chi_{1'}} \gamma_1\xi_1^{3}( \chi_{ {1} 1'};  {z}_{1}, z_{1'}) \bigg] \mathcal{P}_{1}( { {{\bf 
n}}}_1 \cdot  {{\bf n}}_{ {1}1'}) \nonumber \\ 
&+&\bigg[ \left(1-\frac{1}{7}\beta_1\right) \xi_2^{0}( \chi_{ {1} 1'};  {z}_{1}, z_{1'}) 
+ \left(2\frac{\beta_1 \alpha_1}{\chi_{1'}\chi_1}-\gamma_1 \right) \xi_2^{2}( \chi_{ {1} 1'}; {z}_{1}, 
z_{1'}) \bigg] \mathcal{P}_{2}( { {{\bf n}}}_1 \cdot  {{\bf n}}_{ {1}1'}) \nonumber \\ 
&+& \frac{3}{5}\beta_1\left(\frac{\alpha_1}{\chi_1}-\frac{2}{\chi_{1'}}\right) \xi_3^{1}( \chi_{ {1} 1'}; 
{z}_{1}, z_{1'}) \mathcal{P}_{3}( { {{\bf n}}}_1 \cdot  {{\bf n}}_{ {1}1'}) 
- \frac{12}{35}\beta_1\xi_4^{0}( \chi_{ {1} 1'}; {z}_{1}, z_{1'}) \mathcal{P}_{4}( { {{\bf n}}}_1 \cdot  {{\bf 
n}}_{ {1}1'})\bigg\}\;. 
\end{eqnarray} 

${\kappa s}$ correlation:
\begin{eqnarray} 
{\xi}_{\kappa s}( {\bf x}_1, {\bf x}_1)&=& b_1^2 \int^{ {\chi}_{1}} {d\chi_{1'}}   \sigma_{1 1'} 
\bigg\{\bigg[ \left(1+\frac{1}{5}\beta_1 \right) \xi_0^{0}( \chi_{1'  1}; z_{1'},  {z}_{1}) 
- \left(\frac{\beta_1 \alpha_1}{\chi_{1}\chi_{1'}}+\gamma_1 \right) \xi_0^{2}( \chi_{1'  1};  z_{1'},   
{z}_{1}) \bigg] \mathcal{P}_{0}( { {{\bf n}}}_1 \cdot  {{\bf n}}_{1'  1}) \nonumber \\ 
&-&  \bigg[ 3 \left(\frac{1}{\chi_{1'}}+\frac{3}{5\,\chi_{1'}}\beta_1 + \frac{1}{5}\frac{\beta_1 \alpha_1} 
{\chi_{1}}\right) \xi_1^{1}( \chi_{1'  1};  z_{1'},  {z}_{1}) 
- \frac{3}{\chi_{1'}} \gamma_1 \xi_1^{(3)}( \chi_{1' {1}};  z_{1'},  {z}_{1}) \bigg] \mathcal{P}_{1}( { {{\bf 
n}}}_1 \cdot  {{\bf n}}_{1'1}) \nonumber \\ 
&+&\bigg[ \left(1-\frac{1}{7}\beta_1 \right) \xi_2^{0}( \chi_{1'  {1}};  z_{1'},  {z}_{1}) 
 + \left(2\frac{\beta_1 \alpha_1}{\chi_{1'}\chi_{1}}-\gamma_1 \right) \xi_2^{2}( \chi_{1'1};  z_{1'},   
{z}_{1}) \bigg] \mathcal{P}_{2}( { {{\bf n}}}_1 \cdot  {{\bf n}}_{1'1}) \nonumber \\ 
&-& \frac{3}{5}\beta_1\left(\frac{\alpha_1}{\chi_{1}}-\frac{2}{\chi_{1'}}\right) \xi_3^{(1)}( \chi_{1'1};   
z_{1'},  {z}_{1}) \mathcal{P}_{3}( { {{\bf n}}}_1 \cdot  {{\bf n}}_{1'1}) 
- \frac{12}{35}\beta_1\xi_4^{0}( \chi_{ 1'1};  z_{1'}, {z}_{1}) \mathcal{P}_{4}( { {{\bf n}}}_1 \cdot  {{\bf 
n}}_{1'1})\bigg\}\;. 
\end{eqnarray}

${I\kappa} $ correlation:
\begin{eqnarray}
{\xi}_{I\kappa}( {\bf x}_1, {\bf x}_1)&=& b_1^2 \int^{ {\chi}_{1}} {d\chi_{1'}}\int^{ {\chi}_1}{d\chi_{1''}} \;
  \mu_{11'}\sigma_{11'' }\bigg[-  \xi_0^{2}( \chi_{1'1''}; z_{1'}, z_{1''}) \mathcal{P}_{0}( { {{\bf n}}}_1 \cdot  {{\bf n}}_{1'1''}) \nonumber \\
&&~-  \frac{3}{\chi_{1'}} \xi_1^{3}( \chi_{1' 1''}; z_{1'} , z_{1''}) \mathcal{P}_{1}( { {{\bf n}}}_1 \cdot  {{\bf n}}_{1'1''})- \xi_2^{2}( \chi_{1'1''}; z_{1'}, z_{1''}) \mathcal{P}_{2}( { {{\bf n}}}_1 \cdot  {{\bf n}}_{1'1''})\bigg]\;.
\end{eqnarray}

${\kappa I}$ correlation:
\begin{eqnarray}
{\xi}_{\kappa I}( {\bf x}_1, {\bf x}_1)&=& b_1^2 \int^{ {\chi}_{1}} {d\chi_{1'}}\int^{ {\chi}_1}{d\chi_{1''}} \;
 \sigma_{11'}  \mu_{11'' } \bigg[-  \xi_0^{(2)}( \chi_{1'1''}; z_{1'}, z_{1''}) \mathcal{P}_{0}( {{\bf n}}_1 \cdot  {{\bf n}}_{1'1''}) \nonumber \\
&&~+  \frac{3}{\chi_{1}} \xi_1^{(3)}( \chi_{1' 1''}; z_{1'} , z_{1''}) \mathcal{P}_{1}( { {{\bf n}}}_1 \cdot  {{\bf n}}_{1'1''})- \xi_2^{(2)}( \chi_{1'1''}; z_{1'}, z_{1''}) \mathcal{P}_{2}( { {{\bf n}}}_1 \cdot  {{\bf n}}_{1'1''})\bigg]\;.
\end{eqnarray}

Note that $  {{\bf n}}_{1'1''} \neq -  {{\bf n}}_{1''1'}$ $\chi_{1'}=\chi_{1''}$, and $
{{\bf n}}_{ {1}1'}\neq -  {{\bf n}}_{1' {1}}$ unless  $\chi_{1}= {\chi}_{1'}$. This is connected to the fact that we
have already fixed galaxy $1$ and galaxy $2$ {\it a priori}
and by the definition of $ { {{\bf n}}}_{11}$, when we impose the
plane-parallel limit. Indeed the second subscript $1$ is the
second galaxy and not the first one -- i.e. in the flat-sky limit $
{ {{\bf n}}}_{11}=- { {{\bf n}}}_{21}$.

\end{document}